\documentclass[aps,11pt]{revtex4}
\usepackage{dcolumn}
\usepackage{graphicx}
\usepackage{amsmath}
\usepackage{amsfonts}
\usepackage{amssymb}
\usepackage{psfrag}
\usepackage{wrapfig}
\usepackage{subfigure}
\usepackage{makeidx}
\usepackage{bm}
\usepackage{epsf}
\usepackage{hyperref}
\usepackage{color}
\usepackage{cases}
\usepackage{multirow}
\usepackage{float} 
\usepackage{array}
\usepackage{url}
\usepackage{natbib}
\usepackage{makecell} 
\makeatother
\makeatletter

\newcommand{\Rmnum}[1]{\uppercase\expandafter{\romannumeral #1}}

\usepackage{booktabs}

\begin{document} 

\title{Quasinormal modes of type II-perturbation for dilaton-Euler-Heisenberg black holes}

\author{Sheng-Yuan Li$^{1}$\footnote{shengyuanli77@outlook.com}, Yun Soo Myung$^{2}$\footnote{ysmyung@inje.ac.kr}, De-Cheng Zou$^{3}$\footnote{Corresponding author:dczou@jxnu.edu.cn} and Yu-Xuan Wang$^{3}$\footnote{yx.wang.researchgate@outlook.com} }

\address{
$^{1}$Center for Gravitation and Cosmology, College of Physical Science and Technology, Yangzhou University, Yangzhou 225009, China\\
$^{2}$Center for Quantum Spacetime, Sogang University, Seoul 04107, Republic of Korea\\
$^{3}$School of Physics, Jiangxi Normal University, Nanchang 330022, China}
\date{\today}

\begin{abstract}
{
\indent

We study the quasinormal modes (QNMs) of Type II perturbations for dilaton-Euler-Heisenberg (dEH) black holes. These perturbations consist of coupled even-parity gravitational and dilaton perturbations together with odd-parity electromagnetic perturbations. The background is described by mass \(M\), magnetic charge \(Q_m\), and dilaton coupling difference \(\zeta=\alpha-\beta\) to the Euler-Heisenberg term. To find the QNM frequencies, we need to find the parameter space of \((\zeta, Q_m/M)\) that is free from vector ghosts.
For the \(\ell=1\) mode, the radiative block couples the dilaton and axial electromagnetic amplitudes, whereas the \(\ell=2\) mode also contains metric amplitudes, so the frequencies are obtained from matrix-valued boundary-value problems. We calculate QNM frequencies for the fundamental (\(n=0\)) branch with direct integration and the Chebyshev pseudospectral method, and compare the two computations wherever their charge intervals overlap. We find that the \(\ell=1,2\) modes in the \(n=0\) branch remain damped over the ghost-free parameter domain, supporting the stability of dEH black holes against Type II perturbations. The Type I sector will be presented elsewhere; no claim is made here about that sector, higher multipoles, or overtones.
}
\end{abstract}


\maketitle

\section{Introduction}
\label{intro}

As a nonlinear extension of quantum electrodynamics (QED), the
Euler-Heisenberg (EH) theory has shown an effective description of vacuum
polarization in the strong-field regime~\cite{Heisenberg:1936nmg}.  The
quantum vacuum then behaved as a polarizable medium, so that virtual charged
degrees of freedom generated nonlinear polarization and magnetization~\cite{Obukhov:2002xa}. Electrodynamic self-interaction could 
change both the near-horizon geometry and the propagation of perturbations,
which makes EH black holes a useful setting for testing Einstein-Euler-Heisenberg-dilaton (EEHd) theory.

The  magnetically charged EH black-hole solution was firstly obtained in
Ref.~\cite{Yajima:2000kw}, and later works have focused on electric charge,
rotation, regular geometries, and other modified-gravity extensions
~\cite{Ruffini:2013hia,Breton:2019arv,Amaro:2022yew,Guerrero:2020uhn,Nashed:2021ctg}.
Recently motivated by the low-energy limit of string theory and by Lovelock
corrections, Bakopoulos \textit{et al.} have coupled a dilaton to the nonlinear
EH sector and constructed the analytic dilaton-Euler-Heisenberg (dEH) solution used here~\cite{Bakopoulos:2024hah}.
Its geodesics, lensing, shadows, and accretion properties have since been
examined in Refs.~\cite{Yasir:2025npe,Vachher:2024ezs,Huang:2025jfa,Myung:2025zxu,Jiang:2024njc}, and the polar-metric perturbation of dEH black holes was investigated in~\cite{Li:2026gqi},
but the quasinormal mode (QNM) spectrum of the full coupled perturbations in this background has not yet been studied.

It is well-known that QNMs  characterize the linear response of dissipative
systems and determine the ringdown signal after a black-hole perturbation
~\cite{Berti:2007dg,Berti:2009kk}.  The real part of a QNM frequency sets the
oscillation scale, whereas its imaginary part gives the damping or growth
rate, so the sign and magnitude of the latter provide a direct stability
diagnostic.  Since QNM eigenfunctions obey radiative rather than normalizable
boundary conditions, 
the associated perturbation operator is non-Hermitian~\cite{Nollert:1998ys}, and
QNM spectroscopy can test black-hole parameters, the no-hair hypothesis, and
possible modified-gravity effects~\cite{Berti:2005ys,Berti:2007zu,Isi:2019aib,Cardoso:2017cqb,Cardoso:2016rao,Cardoso:2019rvt,Blazquez-Salcedo:2016enn,Franciolini:2018uyq,Cano:2021myl}.
However,  decoupled single-field formulations are not directly applied  for computing  their QNM frequencies  because the metric, dilaton and nonlinear electromagnetic
perturbations are closely coupled in this background and thus, the numerical methods require a matrix-valued problem.

In this work, we study Type II perturbations around dEH black holes, which consist of coupled even-parity metric and dilaton perturbations together with odd-parity electromagnetic perturbations. The complementary Type I sector will be presented elsewhere. We adopt the Regge-Wheeler gauge~\cite{Regge:1957td} and use the parity convention in which axial harmonics acquire \((-1)^{\ell+1}\) with angular index \(\ell\), whereas polar harmonics acquire \((-1)^\ell\). The aim of this paper is therefore to derive the coupled equations for these Type II perturbations and determine thoroughly whether their fundamental QNMs remain damped over the admissible domain of the dEH parameter. In particular, we obtain the QNMs of the coupled even-parity metric and dilaton perturbations together with the odd-parity electromagnetic perturbations. For this purpose, we employ direct integration and the pseudospectral method as two independent numerical approaches, and find that they are in excellent agreement.
We derive the perturbation equation for \(\ell=0\), which is identical to that in a previous work~\cite{Li:2026gqi}, and thus we skip its discussion here. For \(\ell=1\) and \(\ell=2\), we obtain the corresponding perturbation equations, solve them for the QNMs, and finally determine the dependence of the QNM frequencies on the magnetic charge \(Q_m\) with the other parameters fixed.
We further quantify, by comparing with the reduced formulation of Ref.~\cite{Li:2026gqi}, the spectral shifts produced by the odd-parity electromagnetic perturbations.

This paper is organized as follows. In Sec.~\ref{sec2}, we introduce the action for the EEHs theory and the dEH black hole background by specifying the admissible horizon and ghost-free regions. In Sec.~\ref{sec3}, we derive the coupled equations for the Type II perturbations, namely the polar metric, axial electromagnetic, and dilaton perturbations, in the monopole, dipole, and quadrupole sectors. We also describe the direct integration and pseudospectral methods, including the implementation of the ingoing/outgoing boundary conditions, the numerical integration, and the matching strategy. The Chebyshev discretization of the perturbation equations is employed as a generalized eigenvalue problem for computing the QNM frequencies. In Sec.~\ref{sec:updated-numerical-results}, we present the numerical results and analyze the QNM frequencies of the coupled perturbation theory with particular attention to their behavior. We then demonstrate good agreement between the two independent numerical approaches. We summarize our results in Sec.~\ref{sec6}, with the dimensionless perturbation equations in Appendix A and a comparison of QNM frequencies with those obtained by neglecting electromagnetic perturbations in Appendix B.

\section{Black hole solutions with ghost-free region}
\label{sec2}

Considering  the low-energy limit of string theory and the Lovelock framework,
 the Einstein-Maxwell-dilaton theory with a dilaton coupling  to the nonlinear EH term (EEHd theory)~\cite{Bakopoulos:2024hah}
 is given by 
\begin{eqnarray}
S_{\rm EEHd}=\frac{1}{16\pi}\int  d^4x\sqrt{-g}\Big[R-2\nabla^\mu\phi\nabla_\mu\phi-e^{-2\phi}F^2-f(\phi)\left(2\alpha F^\mu_{~\nu} F^\nu_{~\rho} F^\rho_{~\delta} F^\delta_{~\mu}-\beta F^4\right)\Big],\label{action}
\end{eqnarray}
where \(R\) is the Ricci scalar, \(f(\phi)\) is the dilaton coupling
function, \(F^2=F_{\mu\nu}F^{\mu\nu}\), and
\(F^4=(F^2)^2\), with
\(F_{\mu\nu}=\partial_\mu A_\nu-\partial_\nu A_\mu\).  The constants
\(\alpha\) and \(\beta\) specify the two nonlinear electromagnetic
couplings.  For the nondilatonic truncation with
\(\alpha=\beta=1/2\) and \(\phi=0\), one found 
the Reissner-Nordstr\"om (RN)-like solution~\cite{Liu:2019rib} as
\begin{eqnarray}
&&ds^2_{\rm dRN}=\bar{g}_{\mu\nu}dx^\mu dx^\nu=-f(r) dt^2+\frac{dr^2}{f(r)} +r^2d\Omega^2_2,  \nonumber \\
&& f(r)=1-\frac{2M}{r}+\frac{P^2}{r^2}+\frac{Q^2}{r^2} {}_2F_1\Big[\frac{1}{4},1;\frac{5}{4};-\frac{4P^2}{r^4}\Big], \label{dRN-bh}\\
&& \bar{A}=\bar{v}(r,P,Q)dt+P\cos\theta d\varphi \nonumber
\end{eqnarray}
with \({}_2F_1[\cdots]\) denoting the hypergeometric function.  For large
\(r\), the metric function takes the form 
 \begin{equation} \label{f-series}
  f(r)=1-\frac{2M}{r}+\frac{Q^2+P^2}{r^2}-\frac{4Q^2P^2}{5r^6}+\frac{16Q^2P^4}{9r^{10}}-\frac{64 Q^2P^6}{13 r^{14}}+\cdots,
 \end{equation}
where the first three terms reproduce the dyonic RN black hole
geometry, while the remaining terms encode the nonlinear EH corrections.

Variation of the action~\eqref{action} with respect to \(g_{\mu\nu}\), \(\phi\), and \(A_\mu\)
gives the Einstein, dilaton, and generalized Maxwell equations,
\begin{eqnarray}
&&R_{\mu\nu}-\frac{1}{2}R g_{\mu\nu}=2{\partial _\mu}\phi{\partial_ \nu}\phi-g_{\mu\nu}{\partial }^\mu\phi {\partial }_\mu \phi+2T_{\mu\nu},\label{eqG}\\
&&\Box  \phi +\frac{1}{2}e^{-2\phi}F^2-\frac{df(\phi)}{d\phi}\left(\frac{\alpha}{2} F^\mu_{~\nu} F^\nu_{~\gamma} F^\gamma_{~\delta} F^\delta_{~\mu}-\frac{\beta}{4} F^4\right)=0,\label{eqKG}\\
&&{\partial_ \mu}\Big[\sqrt{-g}\left(e^{-2\phi}F^{\mu\nu}+f(\phi)(4\alpha F^\mu_{~\kappa} F^\kappa_{~\lambda} F^{\nu\lambda}-2\beta F^2 F^{\mu\nu})\right)\Big]=0,\label{eqMaxwell}
\end{eqnarray}
where \(T_{\mu\nu}\) is the energy-momentum tensor,
\begin{eqnarray}
T_{\mu\nu}&=&e^{-2\phi}(F^{\alpha}_\mu F_{\nu \alpha}-\frac{1}{4}g_{\mu\nu }F^2)\nonumber\\
&&+f(\phi)\left ( 4\alpha F^\alpha _\mu F^\beta_\nu F^\eta_\alpha F_{\beta\eta} -\frac{1}{2}\alpha g_{\mu\nu} F^\alpha_\beta F^\beta _\gamma F^\gamma _\delta F^\delta_\alpha -2\beta F^\xi_\mu F_{\nu\xi}F^2+\frac{1}{4}g_{\mu\nu}\beta F^4\right).\nonumber
\end{eqnarray}
Here, the dilaton coupling function is chosen as
\begin{eqnarray}
 f(\phi)=-\left[3\cosh(2\phi)+2\right]
 =-\frac{3}{2}\left(e^{-2\phi}+e^{2\phi}+\frac{4}{3}\right).\label{phi}
\end{eqnarray}
Then, equations~\eqref{eqG}--\eqref{eqMaxwell} may admit the magnetically charged
non-spherically symmetric solution~\cite{Bakopoulos:2024hah}
\begin{eqnarray}
 ds^2&=&-H(r) \,dt^2 + \frac{1}{H(r)} \, dr^2 + R(r)^2 \,( d\theta^2 + \sin^2 \theta \, d\varphi^2),\label{oldmetric}\\
H(r)&=&1-\frac{2 M}{r}-\frac{2\zeta Q_m^4}{r^3(r-Q_m^2/M)^3} \quad
R(r)^2=r\left(r-\frac{Q_m^2}{M}\right)
\end{eqnarray}
with dilaton and gauge potential 
\begin{equation}
\bar{\phi} (r)=-\frac{1}{2}\ln \left(1-\frac{Q_m^2}{M r}\right), \quad A_\varphi= Q_m\cos\theta,\label{fRsolution}
\end{equation}
Here,  \(M\) and \(Q_m\) denote the mass and magnetic charge with 
\(\zeta=\alpha-\beta\).  
For \(\zeta=0\) (\(\alpha=\beta\)), this solution
\eqref{fRsolution} becomes the GMGHS  black hole
~\cite{Gibbons:1987ps,Garfinkle:1990qj} and further reduces to the
Schwarzschild solution for \(Q_m=0\). 
To understand the spacetime geometry better and to perform the perturbation analysis, it is useful to introduce a transformation from the radial coordinate to the physical coordinate. To distinguish the two radial coordinates, let \(\rho\) denote
the old  coordinate in Eqs.~\eqref{oldmetric}--\eqref{fRsolution}. We introduce 
\(r=R(\rho)=\sqrt{\rho(\rho-Q_m^2/M)}\).  Transforming the solution leads to 
the spherically symmetric dEH geometry with dilaton hair as
\begin{eqnarray}
  ds^2_{\rm dEH} &\equiv& \bar{g}_{\mu\nu}dx^\mu dx^\nu = -A(r)dt^2 + \frac{1}{B(r)} dr^2 + r^2 (d\theta^2 + \sin^2 \theta d\varphi^2),\label{metric}\\
A(r)&=&1-\frac{4 M^2}{Q_m^2+\sqrt{Q_m^4+4 M^2 r^2}}-\frac{2 \zeta Q_m^4}{r^6},\nonumber\\
B(r)&=&1
- \frac{Q_{m}^{4} + 4M^{2}r^{2}}{r^2(Q_{m}^{2} + \sqrt{Q_{m}^{4} + 4M^{2}r^{2}})} +\frac{Q_m^4}{4M^2r^2}
- \frac{\zeta Q_{m}^{4}(Q_{m}^{4} + 4M^{2}r^{2})}{2M^2r^{8}},\nonumber\\
\bar{\phi} (r)&=&-\frac{1}{2}\ln \left(\frac{\sqrt{Q_m^4+4 M^2 r^2}-Q_m^2}{\sqrt{Q_m^4+4 M^2 r^2}+Q_m^2}\right)\label{solution}.
\end{eqnarray}
We note that the old coordinate \(\rho\in[Q_m^2/M,+\infty)\) is transformed into the physical coordinate \(r\in[0,+\infty)\) with the transformation preserving the asymptotically flat region and ensuring that \(r\) serves as an areal radius, so that a two sphere has the area $ 4\pi r^2$.

Before we proceed,  it would be better to determine the permissible parameter space. To ensure that the dEH black hole solution is physically meaningful, one must check that it is free from vector ghosts. A ghost-free theory requires the effective kinetic coupling $\mathcal{L}_{\mathcal{F}}$ to remain positive throughout the exterior spacetime.

From the action~\eqref{action}, the  vector-Lagrangian including higher derivative terms takes the form
\begin{equation}
    \mathcal{L}_V = e^{-2\bar{\phi}} \mathcal{F} + f(\bar{\phi}) \left( 2\alpha F^{\mu}_{~\nu}F^{\nu}_{~\rho}F^{\rho}_{~\delta}F^{\delta}_{~\mu} - \beta \mathcal{F}^2 \right)
    \label{L_vec}
\end{equation}
with  $\mathcal{F}=2{Q_m}^2/r^4$. Hereafter, we choose  $M=1$. The necessary condition to avoid  vector ghosts is to impose 
\begin{equation}
    \mathcal{L}_{\mathcal{F}} = \frac{\partial \mathcal{L}}{\partial \mathcal{F}} = e^{-2\bar{\phi}(r)}+ 2\zeta f(\bar{\phi}) \mathcal{F}>0.
    \label{eq:LFdef}
\end{equation}
According to the sign of  $\zeta$, one can distinguish two cases. In case $\zeta<0$, $\mathcal{L_F}$ remains strictly positive, so ghost modes are absent. In case $\zeta>0$, however, ghost modes may arise for certain values of $Q_m$.  To ensure the validity of the QNM  calculation, the working region outside the horizon must remain ghost-free, which means that the perturbation equations remain hyperbolic. From Eq.~\eqref{L_vec} and its first derivative $\mathcal{L}_\mathcal{F}$, it follows that in the outside region $r>r_h$, $\mathcal{L_F}(r)$ is monotonically increasing as  
\begin{equation}
\frac{d\mathcal{L_F}(r)}{dr}|_{r>r_h}>0. 
\end{equation}
This implies that for fixed $\zeta$ and $Q_m$, the smaller $r$ might induce ghost modes. 
Accordingly,  considering 
\begin{equation}
\mathcal{L_F}(r=r_h)>0,
\label{eq:noghost}
\end{equation}
the permissible  $Q_m$ corresponding to the ghost-free boundary for a given $\zeta$ can be determined.
Further, the parameters are chosen such that the horizon exists, $A(r_h)=B(r_h)=0$ with $r_h>0$, ensuring the singularity is not naked. This imposes the condition to find the critical value of $Q_m$ corresponding to the extremal black hole for a given $\zeta$:
\begin{align}
A(r,\zeta,Q_m)|_{\zeta}=0, \quad 
\frac{dA(r,\zeta,Q_m)}{dr}|_{\zeta}=0.
\label{eq:horizonstructure}
\end{align}

Combining the above discussion, the permissible ranges of $Q_m$ as a function of $\zeta$. In the subsequent QMN calculations, our parameter choices will be important because they restrict the permissible ranges. As examples, the critical values corresponding to $\zeta=1$ and $\zeta=-1$ are given by $Q_m^c=0.642549$ and $Q_m^c=0.826114$, respectively.

\begin{figure}[H]
\centering
\includegraphics[width=0.62\linewidth]{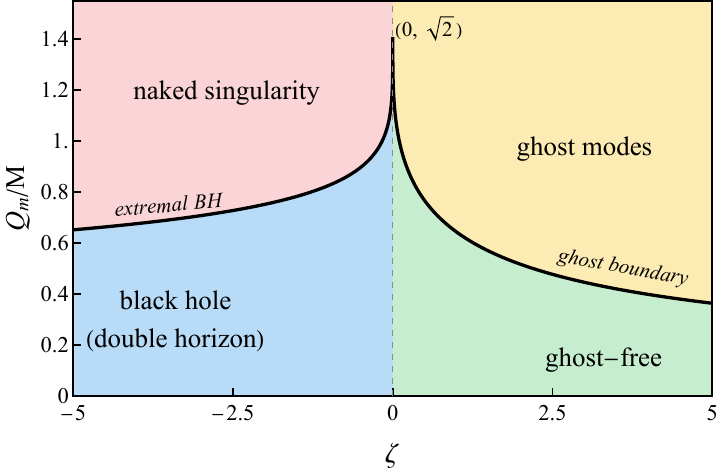}
\caption{Parameter space in the $(\zeta,\, Q_m/M)$ plane.
For $\zeta<0$, the curve corresponds to extremal black holes.
  Below it solutions have two horizons, while above it the
  naked singularity appears.  For $\zeta>0$, the curve represents the onset
  of vector ghost modes.  Below it the parameter space is ghost-free.}
\label{fig:phasediagram}
\end{figure}
Equations \eqref{eq:noghost} and \eqref{eq:horizonstructure} allow us to identify both the vector ghost-free region and the horizon structure of the black hole.  We display  Fig.~\ref{fig:phasediagram} and explain it. For $\zeta<0$, the magnetic charge $Q_m$ is bounded from above by $Q_m^{c}(\zeta)$, beyond which the horizons disappear and a
naked singularity forms. Here, the critical curve coincides with
extremal black holes. 
The junction of two branches at $\zeta=0$ can be read off
from the metric function.  The $\zeta=0$ solution possesses a horizon 
when $r_h=\sqrt{2}\,\sqrt{2-Q_m^{2}}$ is real, implying  $|Q_m|<\sqrt{2}$.
The critical value $Q_m^{c}(\zeta=0)=\sqrt{2}$ corresponds  to the extremal
limit which implies that  the cusp of the critical curve in
Fig.~\ref{fig:phasediagram} is  located at
$(\zeta,Q_m)=(0,\sqrt{2})$.   Note that at $\zeta=0$ the
$\mathcal F$ term vanishes and
Eq.~\eqref{eq:LFdef} implies  $\mathcal{L}_F=e^{-2\phi(r)}>0$, suggesting that no
ghost constraint operates there and the bound is set by horizon
existence alone.
For $\zeta>0$, solutions above the curve (ghost boundary)
acquire ghost modes and lose perturbative unitarity, while those
below the curve remain ghost-free.
The continuity of the critical curve across $\zeta=0$ follows
from the $\zeta\to0$ limits of the two conditions.  Approaching
from the left, the extremality condition
\eqref{eq:horizonstructure} evaluated on the $\zeta\to0^-$
metric  gives
$\frac{dA}{dr}\big|_{r=r_h}=-\sqrt{4-2Q_m^{2}}\,/\,(Q_m^{2}-4)$,
which vanishes precisely at $Q_m=\sqrt{2}$. On the  approaching from the
right ($\zeta\to0^+$), the ghost-free condition \eqref{eq:noghost} at the
horizon gives $\mathcal{L}_F\big|_{r=r_h}=1-Q_m^{2}/2$, whose zero again sits at $Q_m=\sqrt{2}$.  Therefore,  two
branches meet  at the cusp and the
critical value $Q_m^{c}(\zeta)$ becomes continuous for all $\zeta$. Furthermore,  the horizon structure also changes across the diagram. The lower-left region in Fig.~\ref{fig:phasediagram}
possesses two horizons, outer and inner horizons.  As $\zeta$ increases, the inner
horizon shrinks and merges with the singularity at $r=0$, while for $\zeta\ge0$ a single event horizon survives.
Since the curves and regions in the figure are symmetric under $Q_m\to-Q_m$, we display the positive $Q_m$ branch only.  In the subsequent QNM calculations,  $\zeta$ and $Q_m$ will be restricted to this ghost-free region.

\section{Coupled perturbation equations}
\label{sec3}

We now derive the linearized equations governing the perturbations
of the dEH background in Eq.~\eqref{metric}. Let us consider the full perturbation formalism for the spherically symmetric dEH black hole. Following Ref.~\cite{Nomura:2020tpc}, perturbations of EH black holes can be classified into two types according to their parity: I. Odd‑parity metric perturbations couple to even‑parity electromagnetic perturbations. II. Even‑parity metric perturbations couple to odd‑parity electromagnetic perturbations. The coupling types of dEH black holes are similar to those above. Type I is the same as that of dEH black holes, whereas Type II involves an additional dilaton coupling.
In this paper, we focus on Type II perturbations of the dEH black hole.

Let us introduce three  perturbations 
\begin{eqnarray}
g_{\mu\nu}=\bar{g}_{\mu\nu}+\epsilon h_{\mu\nu},\quad \phi=\bar{\phi}+\epsilon\delta\phi,\quad A_\mu=\bar{A}_{\mu}+\epsilon\delta A_\mu
\end{eqnarray}
where $\epsilon$ is an arbitrarily small bookkeeping parameter.  The barred quantities correspond to the dEH background, while \(h_{\mu\nu}\), \(\delta\phi\), and \(\delta A_\mu\)  denote the first-order perturbations. Static and spherically symmetric black holes permit a Fourier
decomposition in time for scalar, tensor, and vector.  With the convention \(e^{-i\omega t}\), the scalar
perturbation is given by
\begin{equation}
\delta\phi=\sum_{l,\mathfrak{m}}\int d\omega\,e^{-i\omega t} \frac{\phi_1(r)}{r} Y_{l}^{\mathfrak{m}}(\theta,\varphi).
\label{eq:scpert}
\end{equation}
Imposing the Regge-Wheeler (RW) gauge~\cite{Regge:1957td}, we expand the polar metric perturbation in tensor spherical harmonics.  In the RW gauge, the polar sector of the metric perturbation is gauge-fixed and is  described by four  modes, $H_0(r)$, $H_1(r)$, $H_2(r)$, and
$K(r)$ as
\begin{align}
&h_{\mu\nu} = \sum_{l,\mathfrak{m}}\int d\omega\,e^{-i\omega t}Y_{l}^{\mathfrak{m}}(\theta,\varphi) \times
\begin{bmatrix}
A(r)H_0(r) & H_1(r) & 0 & 0 \\
H_1(r) & \frac {H_2(r)}{B(r)} & 0 & 0 \\
0 & 0 & r^2 K(r) & 0 \\
0 & 0 & 0 & r^2\sin^2 \theta  K(r)
\end{bmatrix},\label{eq:polarmetricpert} &
\end{align}
The axial vector takes the form with one mode $u_4$
\begin{flalign}
\delta A_{\mu} &= \sum_{l,\mathfrak{m}}\int d\omega\,e^{-i\omega t}  \left(0,0,-\frac{u_{4}(r)\partial_{\varphi}Y_{l}^{\mathfrak{m}}(\theta,\varphi)}{\sin\theta},\,u_{4}(r)\sin\theta\,\partial_{\theta}Y_{l}^{\mathfrak{m}}(\theta,\varphi)\right).\label{eq:axialEMpert}
\end{flalign}
Because the background is spherically symmetric, the radial equations and
their spectrum are independent of index \(\mathfrak{m}\).  We therefore set \(\mathfrak{m}=0\) after the
harmonic projection without loss of generality~\cite{Regge:1957td}.
Hereafter, we classify perturbed equations according to angular index $\ell$. 

\subsection{$\ell=0$ mode: dilaton}
At $\ell=0$ the polar metric and axial electromagnetic perturbations are eliminated from the dynamics, leaving a master equation for
the dilaton.  A second-order differential equation for the dilaton perturbation~\eqref{eq:scpert} is obtained. Introducing  the tortoise coordinate $r_* = \int \frac{1}{\sqrt{A B}} \, dr$, this equation takes the Schr\"odinger form
\begin{eqnarray}
   \frac{d^{2}\Psi_d(r_{*})}{{dr_{*}}^{2}}+[\omega^{2}-V_{d}(r)]\Psi_d(r_{*})=0,
   \label{wavefunc-1}
\end{eqnarray}
\begin{eqnarray}
V_{d}(r) &&=\frac{B A' \left(3 r^2 \bar{\phi} '^2+1\right)}{2 r}-\frac{A e^{-2 \bar{\phi} }}{2 r^8} \Big\{-r^7 e^{2 \bar{\phi}} \Big[B' \left(1-r^2 \bar{\phi} '^2\right)+2 r B \left(r^2 \bar{\phi} '^2-2\right) \bar{\phi} '^2\Big] \nonumber \\
&&+2 r^4 Q_{m}^2 \left(r \bar{\phi} '-2\right)+6 \zeta  Q_{m}^4 \Big[r \left(e^{4 \bar{\phi} }-1\right) \bar{\phi} '+2 \left(e^{4 \bar{\phi} }+1\right)\Big]\Big\}.\label{Veffl=0}
\end{eqnarray}
Here, a prime denotes \(d/dr\). The $s(\ell=0)$-mode carries no radiative gravitational or electromagnetic degree of freedom, so its dynamics is governed by the single
effective potential for the dilaton \(V_d\).  The $\ell=0$ dilaton equation \eqref{wavefunc-1} with the potential \eqref{Veffl=0} and its QNMs were already found in ref.~\cite{Li:2026gqi}.  Therefore, we will not consider  further  this case.

\subsection{$\ell=1$ mode: dilaton-electromagnetic sector}
Since gravitational dipole radiation is forbidden by conservation laws, physical degrees of freedom (DOF) in this sector are carried primarily by dilaton and electromagnetic dipole, while the $\ell=1$ metric components appear as constrained responses.  Metric perturbations with $\ell=1$ describe only an infinitesimal translation of the black hole and are therefore pure gauge \cite{Regge:1957td, Martel:2005ir}.  Once they are removed, the dynamics reside in the dilaton and axial electromagnetic dipole. 
To show the coupling structure up, we introduce three combinations that
occur repeatedly in the radial equations as 
\begin{align}
 \mathcal C_r(r)&=4 e^{2\bar{\phi}(r)}+3 e^{4\bar{\phi}(r)}+3,
 \label{eq:Cr-def}\\
 \mathcal D_r(r)&=r^4-2\zeta Q_m^2\mathcal C_r(r),
 \label{eq:Dr-def}\\
 \mathcal E_r(r)&=r^4+6\zeta Q_m^2
 \left[ e^{4\bar{\phi}(r)}-1\right].
 \label{eq:Er-def}
\end{align}
These definitions are abbreviations only because  they do not introduce additional
physical modes.

On the other hand, the dilaton perturbation $\phi_1(r)$ satisfies
\begin{align}
0={}&\phi_1''
+\frac{BA'+AB'}{2AB}\phi_1'
\nonumber\\
&+\Bigg\{\frac{\omega^2}{AB}
-\frac{1}{2r^8B}\Bigg[
 \frac{r^7BA'}{A}-8r^6A+r^6(rB'+8B+4)
\nonumber\\
&\hspace{3.2cm}
 +4r^4Q_m^2 e^{-2\bar{\phi}}
 -12\zeta Q_m^4 e^{-2\bar{\phi}}
 \left( e^{4\bar{\phi}}+1\right)
 \Bigg]\Bigg\}\phi_1
\nonumber\\
&+\frac{4Q_m e^{-2\bar{\phi}}\mathcal E_r}{r^7B}\,u_4
-\frac{4Q_m e^{-2\bar{\phi}}\bar{\phi}'}{r^5}
 \left(2\zeta Q_m^2\mathcal C_r-r^4\right)u_4'.
\label{eq:l1-scal2}
\end{align}
Furthermore, the electromagnetic perturbation $u_4(r)$ satisfies
\begin{align}
0={}&u_4''
+\frac12\Bigg\{\frac{A'}{A}+\frac{B'}{B}
+\frac{-8\zeta Q_m^2
 \left[3r(e^{4\bar{\phi}}-1)\bar{\phi}'-2\mathcal C_r\right]
 -4r^5\bar{\phi}'}{r\mathcal D_r}\Bigg\}u_4'
\nonumber\\
&+\left[\frac{\omega^2}{AB}
+\frac{12\zeta Q_m^2\mathcal C_r-2r^4}
 {Br^2\mathcal D_r}\right]u_4
+\frac{2Q_m\mathcal E_r}{r^3B\mathcal D_r}\phi_1.
\label{eq:l1-maxeq2}
\end{align}
We emphasize that Eqs.~\eqref{eq:l1-scal2} and
\eqref{eq:l1-maxeq2} are coupled because  the dilaton amplitude is
driven by $u_4$ and $u_4'$, while the electromagnetic amplitude
is driven back by $\phi_1$, through terms that all involve the
magnetic charge $Q_m$ and the combinations
$\mathcal{C}_r$, $\mathcal{D}_r$, $\mathcal{E}_r$.  This describes  the imprint of the
magnetically charged background, which links the even-parity
dilaton to the odd-parity electromagnetic dipole.  For
$Q_m=0$,  two equations become  independent form.  The
QNM spectrum will  therefore  be computed from the coupled
system as a whole,  rather than from two separate
equations.

\subsection{$\ell=2$ mode: gravitational-dilaton-electromagnetic sector}

We finally turn to the radiative sector $\ell\ge2$, where the polar gravitational, axial electromagnetic, and dilaton perturbations are all included.  The polar metric sector is described by four modes  $(H_0, H_1, H_2, K)$, where the $(\theta,\phi)$ component of the linearized Einstein equations sets $H_2=H_0$ algebraically, and it is convenient to further trade $H_1$ for $R_h$ via the frequency-domain redefinition as $H_1=\omega R_h$.  The remaining metric amplitudes are thus $(K, R_h, H_0)$. The metric equations are the three first-order equations \eqref{eq:l2-eqK}--\eqref{eq:l2-eqH0} together with the Einstein constraint \eqref{eq:l2-eqco}, the latter being algebraic in the metric amplitudes and relating $K$, $R_h$, and $H_0$.  Using \eqref{eq:l2-eqco} to fix $H_0$, the metric sector reduces to the two independent variables $K$ and $R_h$, which obey the evolution equations \eqref{eq:l2-eqK} and \eqref{eq:l2-eqR}.  The remaining evolution equation \eqref{eq:l2-eqH0} then follows from the retained equations and the matter equations \eqref{eq:l2-maxeq23} and \eqref{eq:l2-petscal13}. The two wave equations \eqref{eq:l2-maxeq23} and \eqref{eq:l2-petscal13} govern the electromagnetic and scalar amplitudes $u_4$ and $\phi_1$.  Setting $\zeta\to0$ recovers the perturbation equations of the Einstein-Maxwell-dilaton black hole \cite{Gibbons:1987ps}, providing a checking point for our derivation. In obtaining the expressions below, we use three background identities
\begin{align}
 -rB A'\bar{\phi}'+A\left[rB'\bar{\phi}'+2B\left(2\bar{\phi}'+r\bar{\phi}''\right)\right]&=0,\label{eq:L2-bg-id-phi}\\ 
 r\left(\frac{B}{A}\right)'
 +2\left(\frac{B}{A}-1\right)&=0,\label{eq:L2-bg-id-AB}\\ 
 \bar{\phi}'^2+\frac{B-A}{r^2 B}&=0.\label{eq:L2-bg-id-phip2}
\end{align}
These identities follow by substituting the background solution
\eqref{metric} into the background field equations and hold
for arbitrary $M$, $Q_m$, and $\zeta$. In the perturbation equations below
they will be used to eliminate first and second derivatives of the dilaton $\bar\phi''$ and $\bar{\phi}'^2$. 

We briefly describe four equations for the $\ell=2$ mode.
\paragraph{First-order gravitational equations.}
The three gravitational evolution equations read off as 
\begin{align}
0={}&K'-\left(\frac{A'}{2A}-\frac1r\right)K-\frac{H_0}{r}
+\frac{2\bar\phi'}{r}\phi_1
\nonumber\\
&+\frac{ i R_he^{-2\bar\phi}}{2r^8}
\Big\{2r^6 e^{2\bar\phi}(A-2B)
-r^6e^{2\bar\phi}[2rB'+\ell^2+\ell-2]
\nonumber\\
&\hspace{4.2cm}-2r^4Q_m^2+2\zeta Q_m^4\mathcal C_r\Big\},
\label{eq:l2-eqK}\\
0={}&R_h'+\frac12\left(\frac{A'}A+\frac{B'}B\right)R_h
+\frac{ i}{B}(H_0+K)
+\frac{4 i Q_m e^{-2\bar\phi}(2\zeta Q_m^2\mathcal C_r-r^4)}
 {r^6B}u_4,
\label{eq:l2-eqR}\\
0={}&H_0'-\left(\frac1r-\frac{A'}A\right)H_0
-\left(\frac{A'}{2A}-\frac1r\right)K
\nonumber\\
&+ i R_h\Bigg[\frac{1}{2r^8}\Big(
2r^6A-2r^7B'-4r^6B-(\ell^2+\ell)r^6+2r^6
\nonumber\\
&\hspace{3.1cm}
+8\zeta Q_m^4-2r^4Q_m^2 e^{-2\bar\phi}
+6\zeta Q_m^4 e^{-2\bar\phi}+6\zeta Q_m^4 e^{2\bar\phi}
\Big)+\frac{\omega^2}{A}\Bigg]
\nonumber\\
&-\frac{4Q_m e^{-2\bar\phi}(2\zeta Q_m^2\mathcal C_r-r^4)}{r^6}u_4'
-\frac{2\bar\phi'}r\phi_1.
\label{eq:l2-eqH0}
\end{align}

\paragraph{Einstein constraint.}
The $(r,r)$-component of the linearized Einstein equation is imposed as a constraint as 
\begin{align}
0={}&- i R_h\Bigg\{
\frac{rBA'}2\Big[2r^6A e^{2\bar\phi}
-r^6 e^{2\bar\phi}(2rB'+\ell^2+\ell-2)
\nonumber\\
&\hspace{3.8cm}
-4r^6B e^{2\bar\phi}-2r^4Q_m^2+2\zeta Q_m^4\mathcal C_r\Big]
+2r^8\omega^2B e^{2\bar\phi}\Bigg\}
\nonumber\\
&+H_0\Big\{3r^8B e^{2\bar\phi}A'
+A[(\ell^2+\ell-2)r^7 e^{2\bar\phi}+2r^5Q_m^2
-2\zeta r Q_m^4\mathcal C_r]\Big\}
\nonumber\\
&+K\Bigg\{\frac{r^9B e^{2\bar\phi}A'^2}{2A}
-r^8B e^{2\bar\phi}A'
+rA[-(\ell^2+\ell-2)r^6 e^{2\bar\phi}-4r^4Q_m^2
\nonumber\\
&\hspace{4.0cm}+8\zeta Q_m^4\mathcal C_r]
+2r^9\omega^2 e^{2\bar\phi}\Bigg\}
\nonumber\\
&-2\phi_1\Big\{r^8B e^{2\bar\phi}A'\bar\phi'
+2A[r^7B e^{2\bar\phi}\bar\phi'+r^4Q_m^2
+3\zeta Q_m^4( e^{4\bar\phi}-1)]\Big\}
\nonumber\\
&+8r^2ABQ_m\mathcal D_r u_4'
-4r^8AB e^{2\bar\phi}\bar\phi'\phi_1'
+4\ell(\ell+1)rAQ_m\mathcal D_r u_4.
\label{eq:l2-eqco}
\end{align}

\paragraph{Electromagnetic equation.}
The equation for the electromagnetic amplitude $u_4$ takes the form
\begin{align}
0={}&u_4''+\frac12\Bigg\{\frac{A'}A+\frac{B'}B
-\frac{4\{2\zeta Q_m^2[3r( e^{4\bar\phi}-1)\bar\phi'-2\mathcal C_r]
+r^5\bar\phi'\}}{r\mathcal D_r}\Bigg\}u_4'
\nonumber\\
&+\left[\frac{\omega^2}{AB}
-\frac{\ell(\ell+1)(r^4-6\zeta Q_m^2\mathcal C_r)}
 {Br^2\mathcal D_r}\right]u_4
\nonumber\\
&+\frac{Q_m(r^4-6\zeta Q_m^2\mathcal C_r)}{r^2B\mathcal D_r}K
+\frac{2Q_m\mathcal E_r}{r^3B\mathcal D_r}\phi_1.
\label{eq:l2-maxeq23}
\end{align}

\paragraph{Dilaton equation.}
Finally, the equation for the dilaton  $\phi_1$ is given by 
\begin{align}
0={}&\phi_1''+\frac12\left(\frac{A'}A+\frac{B'}B\right)\phi_1'
\nonumber\\
&+\Bigg\{\frac{\omega^2}{AB}
-\frac{1}{2r^8B}\Bigg[
r^6\left(\frac{rBA'}A+rB'
+2[4r^2B\bar\phi'^2+\ell^2+\ell]\right)
\nonumber\\
&\hspace{3.5cm}
+4r^4Q_m^2 e^{-2\bar\phi}
-12\zeta Q_m^4 e^{-2\bar\phi}( e^{4\bar\phi}+1)
\Bigg]\Bigg\}\phi_1
\nonumber\\
&-\frac{2Q_m^2 e^{-2\bar\phi}\mathcal E_r}{r^7B}K
+\frac{2\ell(\ell+1)Q_m e^{-2\bar\phi}\mathcal E_r}{r^7B}u_4
\nonumber\\
&-\frac{4Q_m e^{-2\bar\phi}\bar\phi'(2\zeta Q_m^2\mathcal C_r-r^4)}
 {r^5}u_4'.
\label{eq:l2-petscal13}
\end{align}

The above structure of the $\ell\ge2$ system differs qualitatively from the dipole sector: the metric amplitudes $K$, $R_h$, and $H_0$ become dynamical and couple to both matter amplitudes, so the QNM  spectrum is organized in hybrid gravitational-electromagnetic-dilaton families rather than in separate towers. The constraint \eqref{eq:l2-eqco} supplies the fourth Einstein
relation, fixing the metric combination left undetermined by the evolution equations. Its consistency with \eqref{eq:l2-eqK}--\eqref{eq:l2-eqH0} is readily verified and, as noted above, omitting the constraint admits spurious eigenvalues. We identify the three fundamental branches by continuation from the small-charge regime, where their gravitational, dilaton, and electromagnetic ancestry is unambiguous.  We then track both eigenfrequencies and eigenvectors as the magnetic charge $Q_m$ increases.

\section{Numerical methods}
\label{sec4}

\subsection{Direct integration method}
\label{subsubsec4-A-1-new}
Direct integration and pseudospectral methods treat the same boundary
value problem in complementary ways.  Direct integration is local and makes
the ingoing/outgoing solutions explicit, whereas the pseudospectral method is
global and returns many candidate eigenvalues at once.  Agreement between
them indicates therefore a stronger validation than convergence within either
method alone~\cite{Berti:2009kk, Konoplya:2011qq}.

For direct integration, the radial equations are written as a first-order
matrix system after eliminating redundant variables.  In the
$\ell=1$ sector, for example, we use
\(\boldsymbol\Psi_1^{\rm DI}=(\phi_1,\phi_1',u_4,u_4')^{\mathsf T}\), while the
$\ell=2$ system is reduced analogously after retaining its Einstein constraint.
For \(l=2\) we integrate
\(\boldsymbol\Psi_2^{\rm DI}=(K,R_h,\phi_1,\phi_1',u_4,u_4')^{\mathsf T}\). Here, 
\(H_0\) is reconstructed algebraically from the \(rr\) constraint and the
omitted \(H_0'\) equation is checked as a residual.
The resulting system can be written in the compact matrix form
\begin{eqnarray}
\frac{d}{dr}\boldsymbol\Psi+\boldsymbol V(r,\omega)\boldsymbol\Psi=0,
\end{eqnarray}
where \(\boldsymbol V(r,\omega)\) is assembled from the physical-radius equations
\eqref{eq:l1-scal2}--\eqref{eq:l1-maxeq2} (and from the corresponding
$\ell=2$ equations). 

We solve the coupled ordinary differential equations by imposing the
appropriate boundary conditions at the event horizon \(r\to r_h\) and at
spatial infinity \(r\to\infty\).
At the future horizon the physical solution is purely ingoing, while at
spatial infinity it is purely outgoing~\cite{Chandrasekhar:1975zza}.  Thus, we have 
\begin{equation}
\boldsymbol\Psi\propto
\begin{cases}
e^{-i\omega r_*},& r\rightarrow r_h,\\
e^{+i\omega r_*},& r\rightarrow\infty ,
\end{cases}
\label{eq:di-boundary-conditions}
\end{equation}
where the tortoise coordinate $r_*$  is defined by
\begin{eqnarray}
\frac{dr_*}{dr}=\frac{1}{\sqrt{AB}}.
\end{eqnarray}
Substitution of Frobenius expansions at the two boundaries~\cite{Leaver:1985ax,Nollert:1993zz,Bender:1999box} gives
\begin{align}
\boldsymbol\Psi_h&=e^{-i\omega r_*}(r-r_h)^{p_h}
\sum_{k=0}^{N_h}\boldsymbol a_k(r-r_h)^k,
\nonumber\\
\boldsymbol\Psi_\infty&=e^{+i\omega r_*}r^{p_\infty}
\sum_{k=0}^{N_\infty}\frac{\boldsymbol b_k}{r^k}.
\label{eq:di-series}
\end{align}
Here \(p_h\) and \(p_\infty\) are the residual Frobenius powers after the
explicit ingoing and outgoing phase factors have been removed.  They do not
duplicate the leading wave exponents already contained in
\(e^{\mp i\omega r_*}\). We carry out Frobenius expansions at the horizon and spatial infinity. The series is kept up to 5th order as the resulting expressions become excessively long for higher orders.

The recurrence relations determine the higher-order vectors
\(\boldsymbol a_k\) and \(\boldsymbol b_k\) in terms of the independent
leading data.  Independent basis solutions are integrated outward and
inward to a common point \(r_m\) (we use \(r_m \sim 4 r_h\)), where their values and derivatives form a
matching matrix \(X(\omega,r_m)\)~\cite{Blazquez-Salcedo:2016enn}.  A QNM is a complex root of
\begin{equation}
 \det X(\omega,r_m)=0 .
 \label{eq:di-matching}
\end{equation}
Stability of the root under changes of \(r_m\), series order, and integration
tolerance is required to distinguish a physical zero from a numerical
minimum.
For reproducibility, each direct-integration data set should record the
Frobenius exponents and leading vectors, the truncation orders
\(N_h,N_\infty\), the matching radius \(r_m\), arithmetic precision, and
integration and root-finding tolerances.  These implementation values are
reported with the numerical data rather than treated as physical parameters.

\subsection{Pseudospectral method}
\label{subsec:pseudospectral-method}

For the pseudospectral calculation, we use the dimensionless (reduced) variables
\(Q=Q_m/M\), \(\eta=\zeta/M^2\), \(\Omega=M\omega\), and \(y=\rho/M\)
introduced in Appendix~\ref{appendix-A}.  The dipolar and quadrupolar
equations are those in
Eqs.~\eqref{eq:l1-scaly1}--\eqref{eq:l1-maxeqy1} and
Eqs.~\eqref{eq:l2-publication-eqKy1}--\eqref{eq:l2-publication-maxeqy1},
respectively.  For each pair \((Q,\eta)\), \(y_h\) is chosen as the outer
event horizon, and the exterior is compactified by introducing a coordinate $x$ 
\begin{equation}
 x=1-\frac{y_h}{y},\qquad
 y=\frac{y_h}{1-x},\qquad 0\leq x<1 ,
 \label{eq:ps-compact-map}
\end{equation}
so that \(x=0\) and \(x=1\) correspond to the horizon and spatial infinity.

With the time dependence \(e^{-i\omega t}\), the ingoing and outgoing QNM
conditions are incorporated analytically by writing each dimensionless
amplitude as~\cite{Pani:2013pma,Maselli:2015tta}
\begin{equation}
 \Psi_b(x)=e^{i\Omega y}
 (1-x)^{-2i\Omega+p_{\infty,b}}
 x^{-i\Omega/(2\hat\kappa)+p_{h,b}}\,
 \widehat\Psi_b(x),
 \label{eq:ps-factorization}
\end{equation}
where
\begin{equation}
 \hat\kappa=
 \frac{3y_h-4}{2y_h^2}
 +\frac{3(y_h-2)}{2y_h(y_h-Q^2)}
\end{equation}
is the dimensionless surface-gravity factor and
\(\widehat\Psi_b\) is regular on the compact interval.  For the
\(\ell=1\) amplitudes \((\Phi_1,U_4)\), one has 
\[
 (p_{\infty,b})=(1,0),\qquad (p_{h,b})=(0,0).
\]
For the \(\ell=2\) amplitudes ordered as
\((K_p,R_h,H_0,\Phi_1,U_4)\), the corresponding powers are given by
\[
 (p_{\infty,b})=(1,-1,0,0,0),\qquad
 (p_{h,b})=(0,-1,-1,0,0).
\]
Thus, the boundary conditions are already contained in
Eq.~\eqref{eq:ps-factorization}, and only the regular functions
\(\widehat\Psi_b\) are discretized.

Each regular amplitude is represented by a global Chebyshev interpolant~\cite{Canuto2006, Fornberg1996}, and
its derivatives at the collocation points are obtained from the barycentric
differentiation matrices \(D\) and \(D^2\)
~\cite{Boyd2001}.  The endpoint treatment is adapted to the two
systems.  For \(\ell=1\), we use the Chebyshev-Lobatto grid
\begin{equation}
 x_j^{\rm L}=\frac{1}{2}\left[1+\cos\left(\frac{j\pi}{N-1}\right)\right],
 \qquad j=0,\ldots,N-1 ,
 \label{eq:ps-lobatto-nodes}
\end{equation}
and replace the endpoint equation rows by the regularized Frobenius limits of
the transformed equations.  For \(\ell=2\), the analytic factors in
Eq.~\eqref{eq:ps-factorization} allow the use of the open
Chebyshev-Gauss grid
\begin{equation}
 x_j^{\rm G}=\frac{1}{2}\left[
 1+\cos\left(\frac{(2j-1)\pi}{2N}\right)\right],
 \qquad j=1,\ldots,N ,
 \label{eq:ps-gauss-nodes}
\end{equation}
which avoids direct evaluation of removable endpoint singularities.  In both
cases, we have
\begin{equation}
 \widehat\Psi_b'(x_i)\simeq\sum_jD_{ij}\widehat\Psi_b(x_j),
 \qquad
 \widehat\Psi_b''(x_i)\simeq\sum_j(D^2)_{ij}\widehat\Psi_b(x_j).
 \label{eq:ps-nodal-derivatives}
\end{equation}

Stacking the nodal values of all amplitudes gives
\begin{equation}
 M_N(\Omega)\boldsymbol v=0 ,
 \label{eq:ps-discrete-system}
\end{equation}
where \(M_N\) has dimension \(2N\times2N\) for \(\ell=1\) and
\(5N\times5N\) for \(\ell=2\).  In the quadrupolar problem, the three
gravitational evolution equations and the two matter equations first form the
square system.  The single row of the \(H_0\) evolution equation at the Gauss
point closest to \(x=1/2\) is then replaced by the \(rr\)-Einstein constraint.
All other evolution rows are retained, and the same constraint is evaluated
independently over the interior grid.  This one-row bordering prescription
enforces the constrained system without introducing an additional dynamical
degree of freedom.

After the common asymptotic factors have been differentiated analytically,
the assembled operator is quadratic in \(\Omega\) as 
\begin{equation}
 M_N(\Omega)=A_0+\Omega A_1+\Omega^2A_2 .
 \label{eq:ps-qep}
\end{equation}
The coefficient matrices are reconstructed as
\begin{equation}
 A_0=M_N(0),\qquad
 A_1=\frac{M_N(1)-M_N(-1)}{2},\qquad
 A_2=\frac{M_N(1)+M_N(-1)}{2}-M_N(0),
 \label{eq:ps-qep-reconstruction}
\end{equation}
and the reconstruction is verified at an independent complex frequency.
Following common row and column equilibration of the three matrices, the QEP
is solved through the companion generalized eigenvalue problem
\begin{equation}
 \begin{pmatrix}
  -A_1&-A_0\\ I&0
 \end{pmatrix}
 \begin{pmatrix}\Omega\boldsymbol v\\ \boldsymbol v\end{pmatrix}
 =
 \Omega
 \begin{pmatrix}
  A_2&0\\0&I
 \end{pmatrix}
 \begin{pmatrix}\Omega\boldsymbol v\\ \boldsymbol v\end{pmatrix},
 \label{eq:ps-linearization}
\end{equation}
as standard for quadratic matrix pencils~\cite{TisseurMeerbergen2001}.

The generalized spectrum also contains higher overtones and
resolution-dependent roots.  At the first charge value, the branches of
interest are identified by their proximity to the corresponding Schwarzschild
fundamental frequencies and by their dominant nodal field blocks.  Along each
charge scan, they are continued when using a frequency predictor together with the
overlap of normalized right eigenvectors.  Thereafter the field-block
fractions enter only as secondary labels and are not interpreted as physical
energy fractions.  Distinct roots are assigned to the gravitational-, scalar-,
and electromagnetic-led branches according to their continuous Schwarzschild
ancestry.

For every selected eigenpair, we evaluate the residual of the original,
unbalanced quadratic pencil,
\begin{equation}
 \epsilon_{\rm QEP}=
 \frac{\left\|
 (A_0+\Omega A_1+\Omega^2A_2)\boldsymbol v
 \right\|_2}
 {\left(
 \|A_0\|_\infty+|\Omega|\|A_1\|_\infty+
 |\Omega|^2\|A_2\|_\infty
 \right)\|\boldsymbol v\|_2}.
 \label{eq:ps-eigen-residual}
\end{equation}
For the production calculations, one uses \(N=60\) for \(\ell=1\) and \(N=48\) for
\(\ell=2\).  Reliability is assessed from the QEP residual, the independent
quadratic-reconstruction check, and, for \(\ell=2\), the interior
Einstein-constraint residual.  Endpoints and rapidly varying portions of the
branches are additionally recomputed at higher \(N\). Hence,  points that fail these
resolution or constraint tests are flagged and excluded from the publication
trajectories. Finally,  the direct-integration comparison in
Sec.~\ref{subsec:updated-di-ps-comparison} provides an independent validation
of the retained fundamental branches.

\section{Numerical results and discussion}
\label{sec:updated-numerical-results}

We now present the fundamental QNM spectrum of the dEH black hole.  The
radial equations are solved with the Chebyshev pseudospectral formulation
described in Sec.~\ref{subsec:pseudospectral-method}, and selected cases are
recomputed by direct integration.  The time dependence is
\(\exp(-i\omega t)\), so that \(\operatorname{Im}\omega<0\) denotes a damped
mode.  We use the dimensionless frequency \(M\omega\) and the dimensionless
magnetic charge $Q_m/M$ throughout this section. For compactness we also write reduced charge \(Q\equiv Q_m/M\) and reduced coupling \(\eta\equiv\zeta/M^2\), and fix \(M=1\) in all scans, so that \(Q=Q_m\) and \(\eta=\zeta\) are two independent parameters.
All frequencies below correspond to the fundamental overtone \(n=0\). 
Fundamental QNMs of charged and dilatonic black holes have been computed
in related theories~\cite{Ferrari:2000ep, Chen:2004zr, Shu:2004fj, Brito:2018hjh, Myung:2018jvi}. The coupled dEH spectrum presented
here extends this line of work to the Euler-Heisenberg sector.

The section is organized as follows.
Sec.~\ref{subsec:updated-data-selection} fixes the parameter coverage and
the mode-selection criteria. Sec.~\ref{subsec:updated-di-ps-comparison}
validates the pseudospectral frequencies against direct integration.
Secs.~\ref{subsec:updated-frequency-curves}
and~\ref{subsec:updated-complex-trajectories} analyze the charge dependence
of the fundamental frequencies and the complex-frequency trajectories.

\subsection{Parameter coverage and mode selection}
\label{subsec:updated-data-selection}

The pseudospectral spectra contain the dilaton and electromagnetic
sectors for \(\ell=1\), and the gravitational, dilaton, and electromagnetic
sectors for \(\ell=2\).  The field labels identify the Schwarzschild ancestor of each branch. At finite charge, every eigenfunction is generally a mixture of the coupled amplitudes. Here G, S, and EM denote gravitational, scalar, and
electromagnetic perturbations, respectively.
Both multipoles are scanned at the twelve couplings
\(\eta=\pm0.5,\pm1,\pm1.5,\pm2,\pm2.5,\pm3\), each scan starting from
\(Q=0.01\).  The \(\ell=1\) scans use a uniform step \(\Delta q=0.01\), while the
\(\ell=2\) scans use the same baseline step, refined to \(\Delta q=0.002\)
for \(\eta>0\) and for \(\eta=-0.5,-1.5\), where the fundamental branches
vary most rapidly.  For each coupling the charge runs up to the critical
value \(Q_{\rm c}(\eta)\) at the limiting curve of the admissible
parameter space of Fig.~\ref{fig:phasediagram}, listed in
Table~\ref{tab:critical-charges}.  For \(\eta<0\), the boundary is set by
extremal black holes, for \(\eta>0\) by the onset of vector ghost modes. Every retained point satisfies the
exterior and kinetic cuts \(A>0\), \(B>0\), and \(\mathcal {L}_\mathcal{F}>0\), so all
displayed spectra lie inside the hyperbolic, ghost-free region of the
background.

\begin{table}[H]
\centering
\caption{Critical magnetic charge
\(Q_{\rm c}=Q_m^{\rm c}/M\) at the limiting curve of
Fig.~\ref{fig:phasediagram} for the twelve couplings \(\eta\) scanned in
this section. For \(\eta<0\), the limiting curve corresponds to extremal
black holes while for \(\eta>0\), it marks the onset of ghost modes.  Each
charge scan runs from \(Q=0.01\) up to \(Q_{\rm c}(\eta)\).}
\label{tab:critical-charges}
\begin{tabular}{@{}cccc@{}}
\toprule
\(\eta\) & \(Q_{\rm c}(\eta)\) & \(\eta\) & \(Q_{\rm c}(\eta)\) \\
\midrule
\(-3\)   & 0.707419 & \(+0.5\) & 0.765874 \\
\(-2.5\) & 0.727383 & \(+1\)   & 0.642549 \\
\(-2\)   & 0.751737 & \(+1.5\) & 0.568353 \\
\(-1.5\) & 0.782898 & \(+2\)   & 0.516286 \\
\(-1\)   & 0.826114 & \(+2.5\) & 0.47683 \\
\(-0.5\) & 0.897167 & \(+3\)   & 0.445473 \\
\bottomrule
\end{tabular}
\end{table}
The first point of every scan is selected from the
Schwarzschild-connected fundamental mode.
This anchor is sharp: at \(Q=0.01\) the computed frequencies already agree
with the tabulated Schwarzschild fundamental values~\cite{Berti:2007dg} at
the \(10^{-3}\) level or better, e.g. \(M\omega=0.37367-0.08896\,i\) for
the gravitational-led \(\ell=2\) branch, \(M\omega=0.48367-0.09676\,i\) for
the dilaton-led \(\ell=2\) branch, and \(M\omega=0.24825-0.09249\,i\) for the
electromagnetic-led \(\ell=1\) branch, which fixes the sector ancestry of the
three families.

\subsection{Independent comparison with direct integration}
\label{subsec:updated-di-ps-comparison}

An independent check is available for \(\eta=+1\) and \(\eta=-1\).  
The comparison intervals are
\(0.01\leq Q\leq0.60\) for \(\eta=+1\) and
\(0.01\leq Q\leq0.80\) for \(\eta=-1\), with the common endpoint applied
separately to each field.
For a component \(X\in\{\operatorname{Re}(M\omega),
\operatorname{Im}(M\omega)\}\), we define
\begin{equation}
 \delta_X(Q)=\frac{|X_{\rm PS}(Q)-X_{\rm DI}(Q)|}
 {|X_{\rm DI}(Q)|}\times100\%,
 \qquad
\Delta_{M\omega}(Q)=|M\omega_{\rm PS}(Q)-M\omega_{\rm DI}(Q)|.
 \label{eq:updated-error-definitions}
\end{equation}
When the frequency component is close to zero, the relative quantity
\(\delta_X\) is ill-conditioned.  We therefore use the dimensionless absolute
complex discrepancy \(\Delta_{M\omega}\), together with the componentwise values  as the
primary diagnostic in that regime.

Tables~\ref{tab:updated-L1-plus}--\ref{tab:updated-L2-minus} compare the two
methods at representative charge values on the common intervals. For every
fundamental branch, they list the DI and PS frequencies side by side,
followed by the componentwise relative differences \(\delta_{\rm Re}\) and
\(\delta_{\rm Im}\) defined in Eq.~\eqref{eq:updated-error-definitions}, so
that both oscillation frequency and damping rate are tested.

\begin{table}[H]
\centering
\caption{
Representative direct-integration (DI) and pseudospectral (PS) frequencies
of the \(\ell=1\), \(\eta=+1\) fundamental branches (\(M=1\), \(n=0\),
\(0.01\leq Q\leq0.60\)). S and EM denote the scalar-led and
electromagnetic-led branches.  The last two columns give the component-wise
relative differences  defined in
Eq.~\eqref{eq:updated-error-definitions}.}
\label{tab:updated-L1-plus}
\begingroup\scriptsize
\resizebox{\linewidth}{!}{%
\begin{tabular}{@{}ccllrr@{}}
\toprule
$Q$ & Sector & $M\omega_{\rm DI}$ & $M\omega_{\rm PS}$ & $\delta_{\rm Re}(\%)$ & $\delta_{\rm Im}(\%)$ \\
\midrule
0.01 & S  & $0.29289738-0.09770899\,\mathrm{i}$ & $0.29295097-0.09766152\,\mathrm{i}$ & $1.829\times10^{-2}$ & $4.859\times10^{-2}$ \\
     & EM & $0.24821829-0.09235912\,\mathrm{i}$ & $0.24824892-0.09248503\,\mathrm{i}$ & $1.234\times10^{-2}$ & $1.363\times10^{-1}$ \\
0.20 & S  & $0.29842865-0.09825367\,\mathrm{i}$ & $0.29846580-0.09819789\,\mathrm{i}$ & $1.245\times10^{-2}$ & $5.677\times10^{-2}$ \\
     & EM & $0.24261839-0.09138816\,\mathrm{i}$ & $0.24261368-0.09152472\,\mathrm{i}$ & $1.940\times10^{-3}$ & $1.494\times10^{-1}$ \\
0.40 & S  & $0.31235165-0.09946887\,\mathrm{i}$ & $0.31234818-0.09941249\,\mathrm{i}$ & $1.112\times10^{-3}$ & $5.669\times10^{-2}$ \\
     & EM & $0.22304397-0.09090691\,\mathrm{i}$ & $0.22288993-0.09101499\,\mathrm{i}$ & $6.906\times10^{-2}$ & $1.189\times10^{-1}$ \\
0.60 & S  & $0.33183101-0.10194554\,\mathrm{i}$ & $0.33179198-0.10191958\,\mathrm{i}$ & $1.176\times10^{-2}$ & $2.547\times10^{-2}$ \\
     & EM & $0.21555154-0.11394512\,\mathrm{i}$ & $0.21505978-0.11432021\,\mathrm{i}$ & $2.281\times10^{-1}$ & $3.292\times10^{-1}$ \\
\bottomrule
\end{tabular}}
\endgroup
\end{table}

\begin{table}[H]
\centering
\caption{
Same layout as Table~\ref{tab:updated-L1-plus}, for \(\eta=-1\) with
\(0.01\leq Q\leq0.80\).  For both signs of \(\eta\), the listed charges lie
within the admissible ranges of Table~\ref{tab:critical-charges}. The
\(\eta=-1\) interval is longer because its extremality boundary lies
farther out than the ghost-free boundary of \(\eta=+1\).  It includes the
point where the signed \(\operatorname{Im}(M\omega)\) of the
electromagnetic-led branch is most negative.}
\label{tab:updated-L1-minus}
\begingroup\scriptsize
\resizebox{\linewidth}{!}{%
\begin{tabular}{@{}ccllrr@{}}
\toprule
$Q$ & Sector & $M\omega_{\rm DI}$ & $M\omega_{\rm PS}$ & $\delta_{\rm Re}(\%)$ & $\delta_{\rm Im}(\%)$ \\
\midrule
0.01 & S  & $0.29289739-0.09770900\,\mathrm{i}$ & $0.29295097-0.09766152\,\mathrm{i}$ & $1.829\times10^{-2}$ & $4.859\times10^{-2}$ \\
     & EM & $0.24823741-0.09236330\,\mathrm{i}$ & $0.24826816-0.09248917\,\mathrm{i}$ & $1.239\times10^{-2}$ & $1.363\times10^{-1}$ \\
0.20 & S  & $0.29883815-0.09835942\,\mathrm{i}$ & $0.29887409-0.09830287\,\mathrm{i}$ & $1.203\times10^{-2}$ & $5.750\times10^{-2}$ \\
     & EM & $0.25038135-0.09297769\,\mathrm{i}$ & $0.25042722-0.09309943\,\mathrm{i}$ & $1.832\times10^{-2}$ & $1.309\times10^{-1}$ \\
0.40 & S  & $0.31717642-0.10067905\,\mathrm{i}$ & $0.31715886-0.10062540\,\mathrm{i}$ & $5.539\times10^{-3}$ & $5.329\times10^{-2}$ \\
     & EM & $0.25910063-0.09522008\,\mathrm{i}$ & $0.25919483-0.09530640\,\mathrm{i}$ & $3.636\times10^{-2}$ & $9.065\times10^{-2}$ \\
0.60 & S  & $0.35172831-0.10611584\,\mathrm{i}$ & $0.35168737-0.10613686\,\mathrm{i}$ & $1.164\times10^{-2}$ & $1.981\times10^{-2}$ \\
     & EM & $0.28108630-0.09951128\,\mathrm{i}$ & $0.28118679-0.09946948\,\mathrm{i}$ & $3.575\times10^{-2}$ & $4.200\times10^{-2}$ \\
0.80 & S  & $0.41810011-0.11343922\,\mathrm{i}$ & $0.41812019-0.11342078\,\mathrm{i}$ & $4.803\times10^{-3}$ & $1.626\times10^{-2}$ \\
     & EM & $0.32814414-0.09666842\,\mathrm{i}$ & $0.32811089-0.09664795\,\mathrm{i}$ & $1.013\times10^{-2}$ & $2.117\times10^{-2}$ \\
\bottomrule
\end{tabular}}
\endgroup
\end{table}

\begin{table}[H]
\centering
\caption{
Representative DI and PS frequencies of the \(\ell=2\), \(\eta=+1\)
fundamental branches (\(M=1\), \(n=0\), \(0.01\leq Q\leq0.60\)). Here G, S, and
EM label the gravitational-, scalar-, and electromagnetic-led branches.
}
\label{tab:updated-L2-plus}
\begingroup\scriptsize
\resizebox{\linewidth}{!}{%
\begin{tabular}{@{}ccllrr@{}}
\toprule
$Q$ & Sector & $M\omega_{\rm DI}$ & $M\omega_{\rm PS}$ & $\delta_{\rm Re}(\%)$ & $\delta_{\rm Im}(\%)$ \\
\midrule
0.01 & G  & $0.37357911-0.08894068\,\mathrm{i}$ & $0.37367427-0.08896260\,\mathrm{i}$ & $2.547\times10^{-2}$ & $2.464\times10^{-2}$ \\
     & S  & $0.48366395-0.09686040\,\mathrm{i}$ & $0.48366816-0.09676035\,\mathrm{i}$ & $8.697\times10^{-4}$ & $1.033\times10^{-1}$ \\
     & EM & $0.45747523-0.09502402\,\mathrm{i}$ & $0.45757861-0.09500423\,\mathrm{i}$ & $2.260\times10^{-2}$ & $2.083\times10^{-2}$ \\
0.20 & G  & $0.37453411-0.08906631\,\mathrm{i}$ & $0.37461932-0.08908754\,\mathrm{i}$ & $2.275\times10^{-2}$ & $2.383\times10^{-2}$ \\
     & S  & $0.49190733-0.09741264\,\mathrm{i}$ & $0.49187080-0.09732106\,\mathrm{i}$ & $7.427\times10^{-3}$ & $9.401\times10^{-2}$ \\
     & EM & $0.45200844-0.09516695\,\mathrm{i}$ & $0.45211632-0.09517566\,\mathrm{i}$ & $2.387\times10^{-2}$ & $9.153\times10^{-3}$ \\
0.40 & G  & $0.37582081-0.09042066\,\mathrm{i}$ & $0.37587089-0.09043854\,\mathrm{i}$ & $1.332\times10^{-2}$ & $1.977\times10^{-2}$ \\
     & S  & $0.51009638-0.09890309\,\mathrm{i}$ & $0.51000222-0.09887649\,\mathrm{i}$ & $1.846\times10^{-2}$ & $2.689\times10^{-2}$ \\
     & EM & $0.44079420-0.09916673\,\mathrm{i}$ & $0.44089961-0.09924302\,\mathrm{i}$ & $2.391\times10^{-2}$ & $7.693\times10^{-2}$ \\
0.60 & G  & $0.37571030-0.10098175\,\mathrm{i}$ & $0.37561810-0.10096016\,\mathrm{i}$ & $2.454\times10^{-2}$ & $2.139\times10^{-2}$ \\
     & S  & $0.53493394-0.10166484\,\mathrm{i}$ & $0.53487569-0.10174585\,\mathrm{i}$ & $1.089\times10^{-2}$ & $7.969\times10^{-2}$ \\
     & EM & $0.43425774-0.10736857\,\mathrm{i}$ & $0.43433704-0.10748233\,\mathrm{i}$ & $1.826\times10^{-2}$ & $1.060\times10^{-1}$ \\
\bottomrule
\end{tabular}}
\endgroup
\end{table}

\begin{table}[H]
\centering
\caption{
Same layout as Table~\ref{tab:updated-L2-plus}, for \(\eta=-1\) with
\(0.01\leq Q\leq0.80\).  The charges again lie within the admissible
\(\eta=-1\) range of Table~\ref{tab:critical-charges}, and the tabulated
branches remain continuous from the weak-charge regime to the
near-extremal end.}
\label{tab:updated-L2-minus}
\begingroup\scriptsize
\resizebox{\linewidth}{!}{%
\begin{tabular}{@{}ccllrr@{}}
\toprule
$Q$ & Sector & $M\omega_{\rm DI}$ & $M\omega_{\rm PS}$ & $\delta_{\rm Re}(\%)$ & $\delta_{\rm Im}(\%)$ \\
\midrule
0.01 & G  & $0.37357911-0.08894068\,\mathrm{i}$ & $0.37367427-0.08896260\,\mathrm{i}$ & $2.547\times10^{-2}$ & $2.464\times10^{-2}$ \\
     & S  & $0.48366397-0.09686040\,\mathrm{i}$ & $0.48366817-0.09676035\,\mathrm{i}$ & $8.697\times10^{-4}$ & $1.033\times10^{-1}$ \\
     & EM & $0.45750380-0.09502421\,\mathrm{i}$ & $0.45760712-0.09500426\,\mathrm{i}$ & $2.258\times10^{-2}$ & $2.100\times10^{-2}$ \\
0.20 & G  & $0.37472915-0.08905389\,\mathrm{i}$ & $0.37481485-0.08907551\,\mathrm{i}$ & $2.287\times10^{-2}$ & $2.427\times10^{-2}$ \\
     & S  & $0.49366592-0.09745470\,\mathrm{i}$ & $0.49362096-0.09736821\,\mathrm{i}$ & $9.107\times10^{-3}$ & $8.875\times10^{-2}$ \\
     & EM & $0.46220989-0.09509686\,\mathrm{i}$ & $0.46229886-0.09504999\,\mathrm{i}$ & $1.925\times10^{-2}$ & $4.928\times10^{-2}$ \\
0.40 & G  & $0.37936532-0.08946217\,\mathrm{i}$ & $0.37942695-0.08948934\,\mathrm{i}$ & $1.624\times10^{-2}$ & $3.037\times10^{-2}$ \\
     & S  & $0.52728711-0.09953004\,\mathrm{i}$ & $0.52721765-0.09958834\,\mathrm{i}$ & $1.317\times10^{-2}$ & $5.858\times10^{-2}$ \\
     & EM & $0.47702302-0.09602635\,\mathrm{i}$ & $0.47704629-0.09593367\,\mathrm{i}$ & $4.878\times10^{-3}$ & $9.652\times10^{-2}$ \\
0.60 & G  & $0.39027805-0.09039949\,\mathrm{i}$ & $0.39032262-0.09045722\,\mathrm{i}$ & $1.142\times10^{-2}$ & $6.386\times10^{-2}$ \\
     & S  & $0.59921823-0.10622467\,\mathrm{i}$ & $0.59924108-0.10614658\,\mathrm{i}$ & $3.814\times10^{-3}$ & $7.351\times10^{-2}$ \\
     & EM & $0.50408290-0.09769786\,\mathrm{i}$ & $0.50399963-0.09766389\,\mathrm{i}$ & $1.652\times10^{-2}$ & $3.477\times10^{-2}$ \\
0.80 & G  & $0.41069684-0.09055380\,\mathrm{i}$ & $0.41079445-0.09065967\,\mathrm{i}$ & $2.377\times10^{-2}$ & $1.169\times10^{-1}$ \\
     & S  & $0.74860630-0.11052149\,\mathrm{i}$ & $0.74858633-0.11051015\,\mathrm{i}$ & $2.668\times10^{-3}$ & $1.025\times10^{-2}$ \\
     & EM & $0.54939418-0.09594816\,\mathrm{i}$ & $0.54941925-0.09600860\,\mathrm{i}$ & $4.563\times10^{-3}$ & $6.300\times10^{-2}$ \\
\bottomrule
\end{tabular}}
\endgroup
\end{table}


Across all common points and sectors, the largest componentwise discrepancy for $\ell=1$ occurs at $Q=0.54$ for $\mathrm{Re}(M\omega)$ and at $Q=0.58$ for $\mathrm{Im}(M\omega)$. For $Q=0.54$, the direct-integration result reads $M\omega_\mathrm{DI}=0.20321170 - 0.11700601\,i$, while the pseudospectral result is $M\omega_\mathrm{PS}=0.20233624-0.11718381\,i$, yielding a relative discrepancy of $0.4308\%$ in $\mathrm{Re}(M\omega)$. For $Q=0.58$, $M\omega_\mathrm{DI}=0.21315428 - 0.11924291\,i$ and $M\omega_\mathrm{PS}=0.21251074-0.11972669\,i$, which gives $0.4057\%$ discrepancy in $\mathrm{Im}(M\omega)$. Both of these peak-discrepancy points lie on the $\ell=1$, $\eta=+1$ electromagnetic-led branch and are not listed in our tables. 
For $\ell=2$, both the maximal relative discrepancy of the $\mathrm{Re}(M\omega)$ and that of the $\mathrm{Im}(M\omega)$ lie on the $\eta=-1$ gravitational-led branch. The real-part maximum $2.547\times 10^{-2}\%$ occurs at $Q=0.01$, while the imaginary-part maximum $0.1169\%$ occurs at $Q=0.6$.
All differences therefore remain at the sub-percent level, and the two methods agree well over the whole common domain. Charge-induced spectral shifts reach tens of percent for the studied modes. The gravitational quadrupole branch at $\eta=+1$ exhibits the weakest spectral shift. Even this shift exceeds the corresponding DI–PS discrepancy by more than an order of magnitude. The frequency curves analyzed above are numerically resolved rather than discretization artifacts.

\subsection{Frequency dependence and method agreement}
\label{subsec:updated-frequency-curves}

\begin{figure}[H]
\centering
\includegraphics[width=0.96\linewidth]{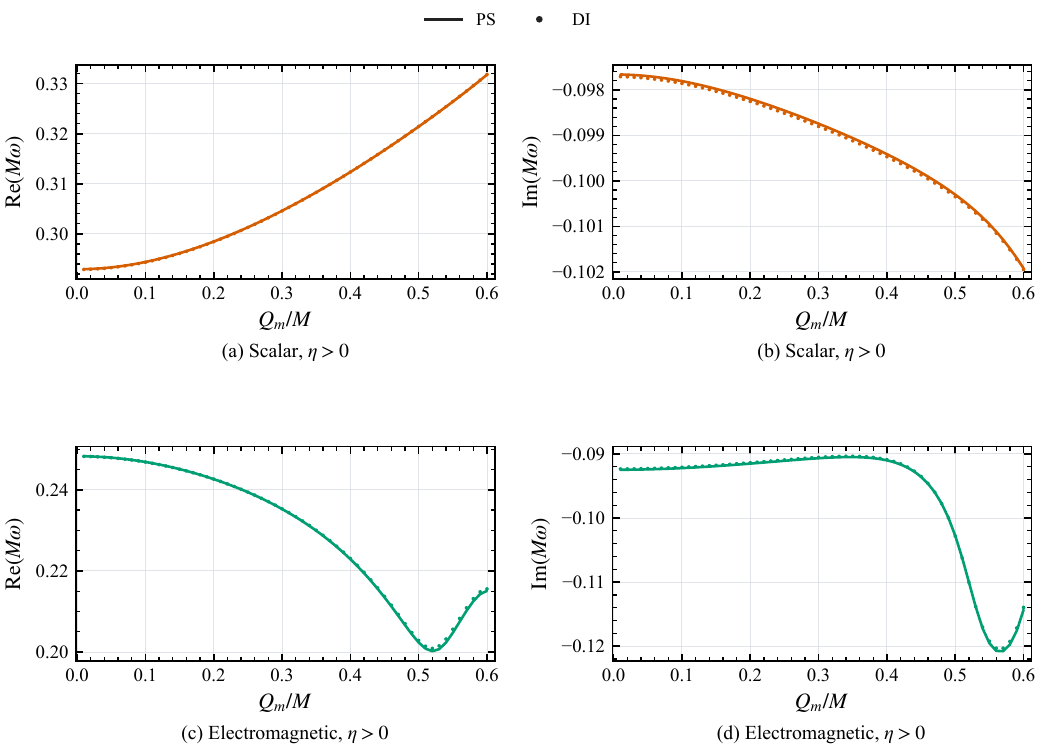}
\caption{
Fundamental \(\ell=1\) frequencies for \(\eta=+1\) (\(M=1\), \(n=0\),
\(0.01\leq Q\leq0.60\)).  Upper and lower rows: scalar-led and
electromagnetic-led branches. Left and right columns:
\(\operatorname{Re}(M\omega)\) and \(\operatorname{Im}(M\omega)\).  Solid
curves are the pseudospectral results,  while filled points represent the direct-integration
values.  The electromagnetic-led branch develops a pronounced extremum of
its signed \(\operatorname{Im}(M\omega)\) near the high-charge end.}
\label{fig:updated-L1-plus}
\end{figure}

\begin{figure}[H]
\centering
\includegraphics[width=0.96\linewidth]{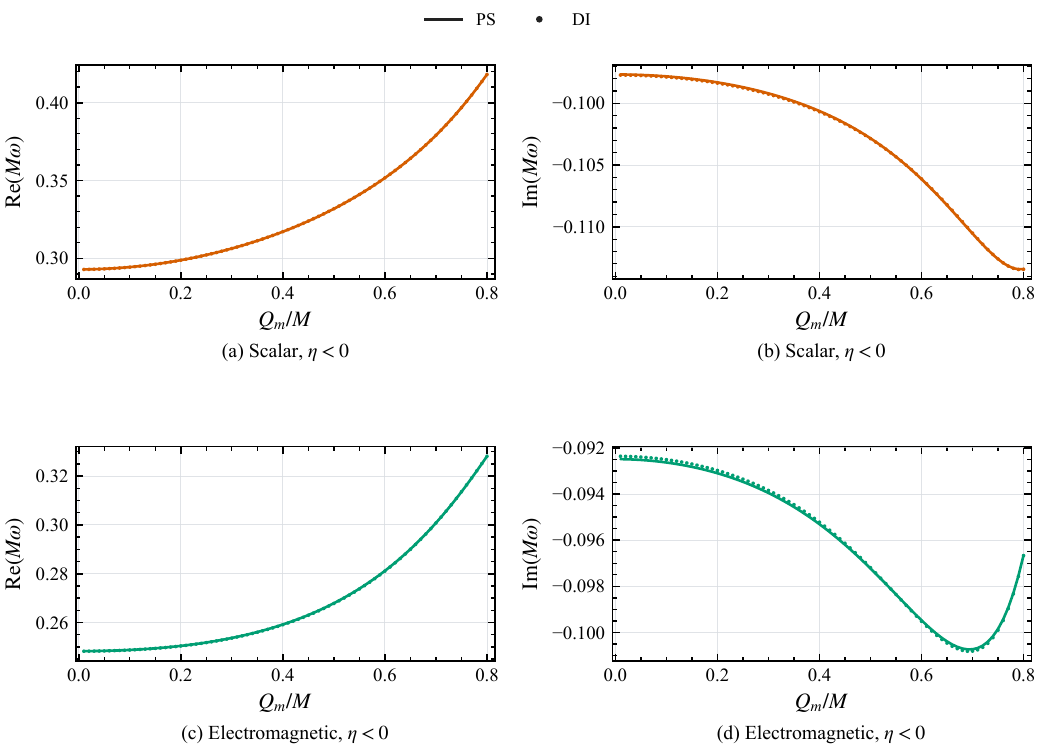}
\caption{
Same as Fig.~\ref{fig:updated-L1-plus}, but for \(\eta=-1\) with
\(0.01\leq Q\leq0.80\).  The scalar-led frequency grows monotonically
with charge, while the signed \(\operatorname{Im}(M\omega)\) of the
electromagnetic-led mode passes through a smooth extremum before the
endpoint.}
\label{fig:updated-L1-minus}
\end{figure}

\begin{figure}[H]
\centering
\includegraphics[width=0.96\linewidth]{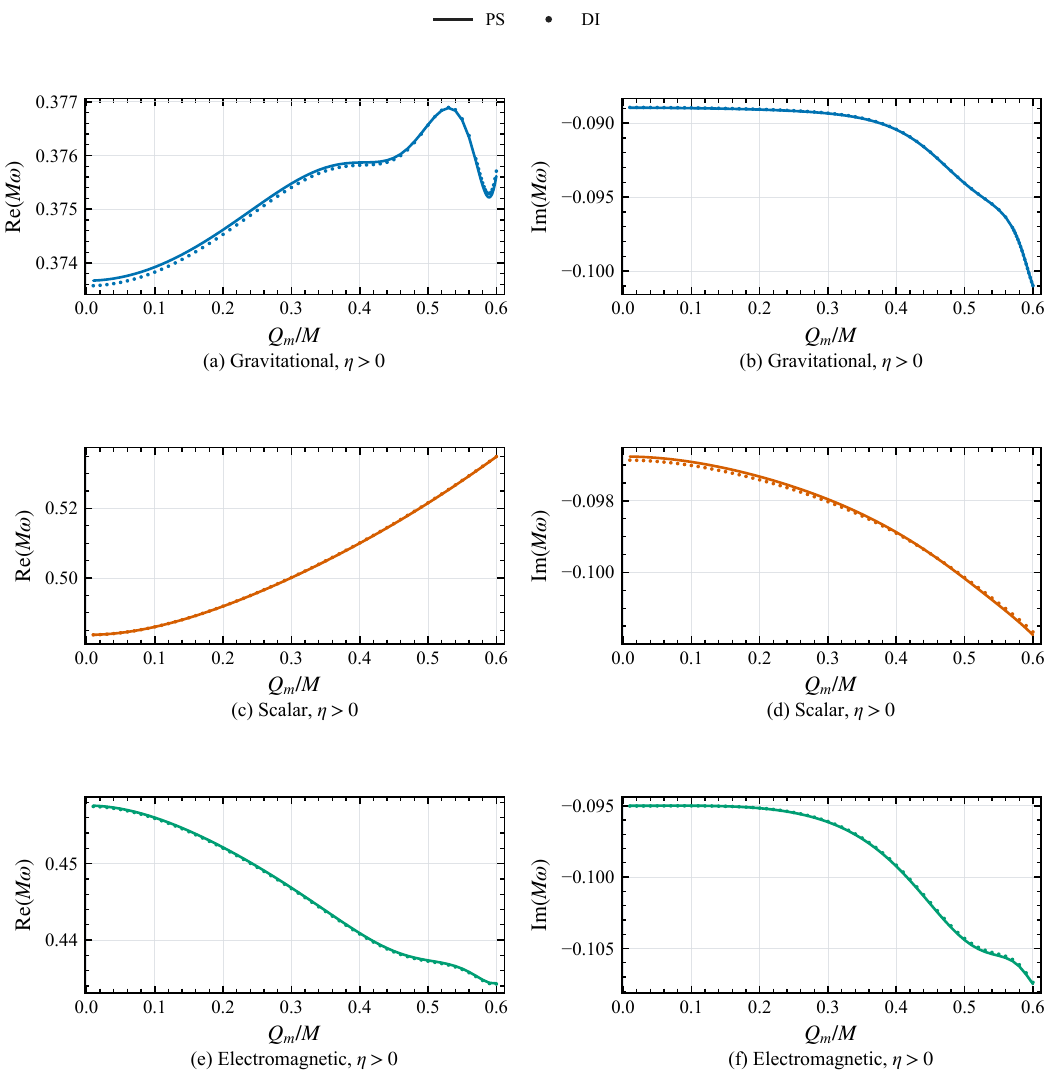}
\caption{
Fundamental \(\ell=2\) frequencies for \(\eta=+1\) (\(M=1\), \(n=0\),
\(0.01\leq Q\leq0.60\)).  Rows: gravitational-led, scalar-led, and
electromagnetic-led branches, while  columns are as in
Fig.~\ref{fig:updated-L1-plus}.  The rapid high-charge variation of the
gravitational-led branch motivates the denser charge grid and the
independent constraint check used near the endpoint.}
\label{fig:updated-L2-plus}
\end{figure}

\begin{figure}[H]
\centering
\includegraphics[width=0.96\linewidth]{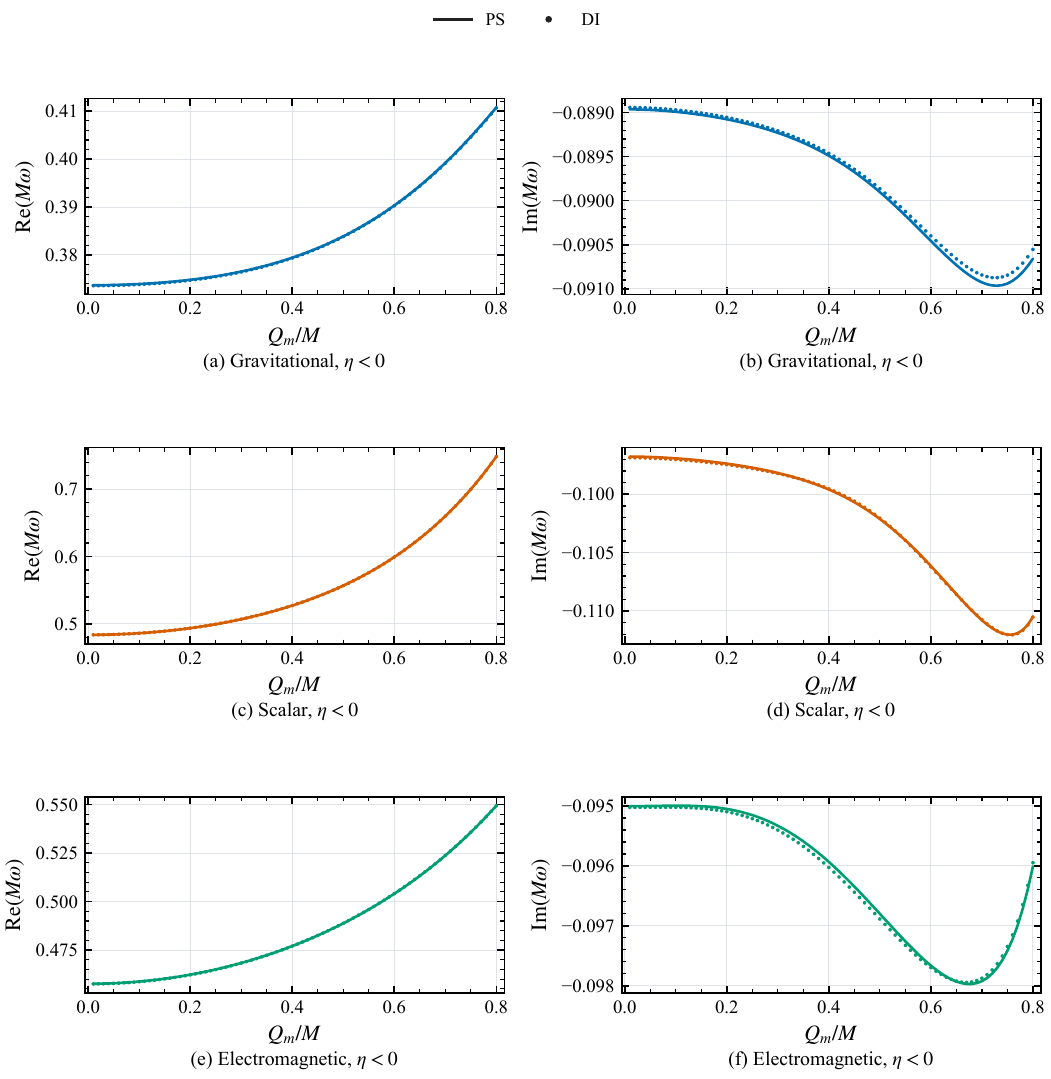}
\caption{
Same as Fig.~\ref{fig:updated-L2-plus}, but for \(\eta=-1\) with
\(0.01\leq Q\leq0.80\).  All three real frequencies increase over the
displayed range, while the imaginary parts vary nonlinearly in charge and
pass through visible extrema.}
\label{fig:updated-L2-minus}
\end{figure}

Figs.~\ref{fig:updated-L1-plus} and~\ref{fig:updated-L1-minus} show
the fundamental \(\ell=1\) frequencies for \(\eta=\pm1\), and
Figs.~\ref{fig:updated-L2-plus} and~\ref{fig:updated-L2-minus} the
\(\ell=2\) frequencies, as functions of magnetic charge \(Q=Q_m/M\). In every panel, the
pseudospectral curves and the direct-integration points track the same
branch over the entire common interval. 
Beyond this internal consistency between the two methods,
Appendix~\ref{app:previous-reduced-comparison} compares the present direct-integration branches with gravitational–dilaton perturbation system results in~\cite{Li:2026gqi} across the common charge range. The two formulations share the neutral limit. It turns out that the retained electromagnetic amplitude shifts the fundamental frequencies by several percent already at \(Q=0.6\) and by up to \(34\%\) at the high-charge endpoints.

The figures also reveal the physical effect of the coupling sign.  For
\(\eta>0\), the \(\ell=1\) scalar-led frequency increases while both
damping rates grow, the electromagnetic-led branch migrating from
\(M\omega=0.24822-0.09236\,i\) at \(Q=0.01\) to
\(M\omega=0.21555-0.11395\,i\) at \(Q=0.60\), a \(13\%\) drop in
oscillation frequency with a \(23\%\) growth in damping.  For
\(\eta<0\), the real parts of both \(\ell=1\) branches increase with
charge, the electromagnetic-led one passing through a smooth minimum of
its signed \(\operatorname{Im}(M\omega)\).  At \(\ell=2\) the
charge-induced shifts are large for the matter-led branches along
\(\eta=-1\), the scalar-led frequency grows by \(55\%\) (from
\(M\omega=0.48366-0.09686\,i\) to \(M\omega=0.74861-0.11052\,i\)
between \(Q=0.01\) and \(Q=0.80\)) and the electromagnetic-led branch
by \(20\%\), while the gravitational-led quadrupole is much stiffer. 
It gains \(10\%\) in frequency with its damping rate within \(2\%\)
of the Schwarzschild value for \(\eta=-1\), but drifts by only
\(0.6\%\) in frequency while its damping rate grows by \(13\%\) for
\(\eta=+1\).

These trends follow from the sign with which the EH term
enters the background functions and the off-diagonal couplings. The piece
\(-2\zeta Q_m^4/r^6\) of \(A(r)\) is concentrated near the horizon,
and reversing its sign redistributes its influence between the
oscillation and damping channels. Also, a quantitative attribution would
require further research of the coupled system, which we
leave to future work.  The bends visible in several curves reflect the
simultaneous deformation of the effective propagation potential and of
the inter-field energy transfer.

\begin{figure}[H]
\centering
\includegraphics[width=0.94\linewidth]{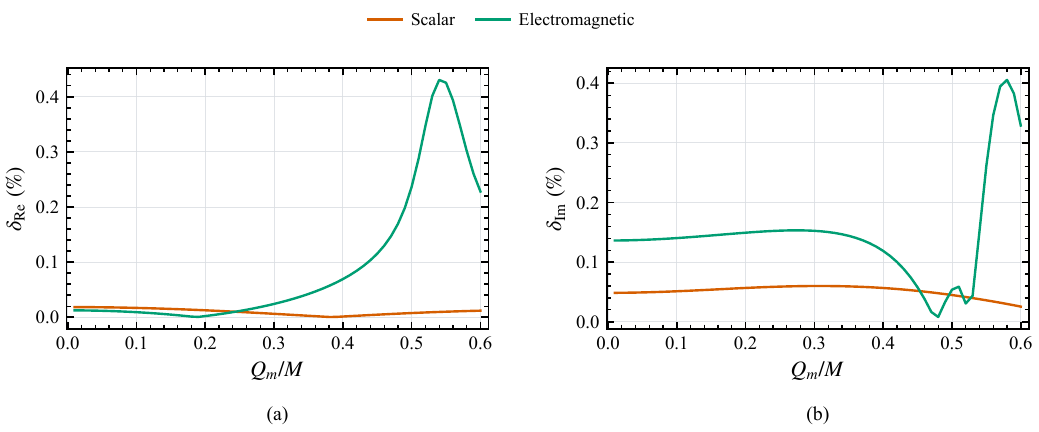}
\caption{
Componentwise DI-PS relative differences for \(\ell=1\), \(\eta=+1\)
(\(M=1\), \(n=0\), \(0.01\leq Q\leq0.60\)). Both sectors share each
panel, with the real- and imaginary-part differences separated.  The
differences are largest on the electromagnetic-led branch  toward the
upper end of the interval.}
\label{fig:updated-L1-errors-plus}
\end{figure}

\begin{figure}[H]
\centering
\includegraphics[width=0.94\linewidth]{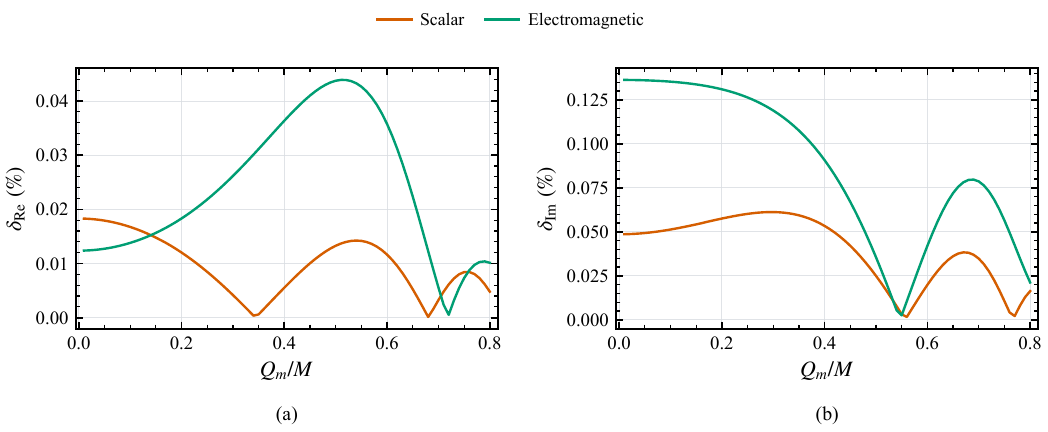}
\caption{
Same as Fig.~\ref{fig:updated-L1-errors-plus}, but for \(\eta=-1\) with
\(0.01\leq Q\leq0.80\).  Both sectors remain sub-percent throughout,
including the extremum region of the electromagnetic-led signed
\(\operatorname{Im}(M\omega)\).}
\label{fig:updated-L1-errors-minus}
\end{figure}

\begin{figure}[H]
\centering
\includegraphics[width=0.94\linewidth]{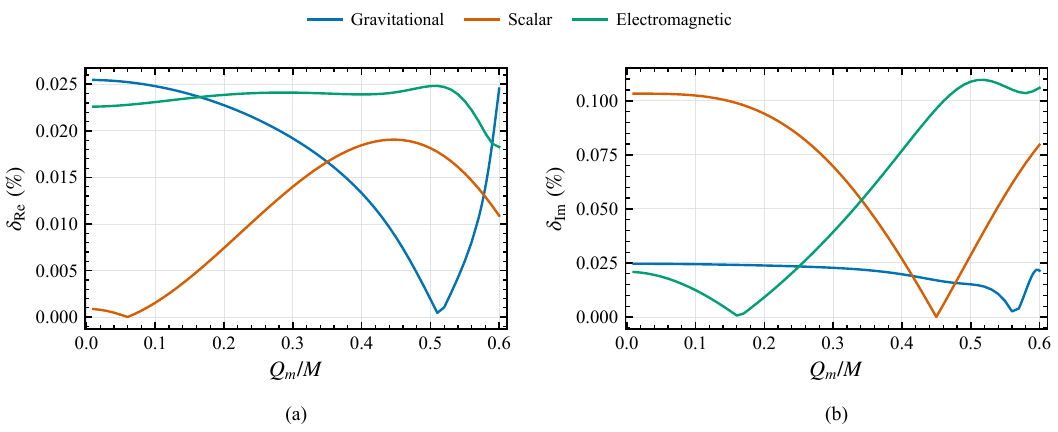}
\caption{
Componentwise DI-PS relative differences for \(\ell=2\), \(\eta=+1\)
(\(M=1\), \(n=0\), \(0.01\leq Q\leq0.60\)). All three sectors share each
panel and remain sub-percent, supporting a consistent solution of the
constrained \(\ell=2\) system.}
\label{fig:updated-L2-errors-plus}
\end{figure}

\begin{figure}[H]
\centering
\includegraphics[width=0.94\linewidth]{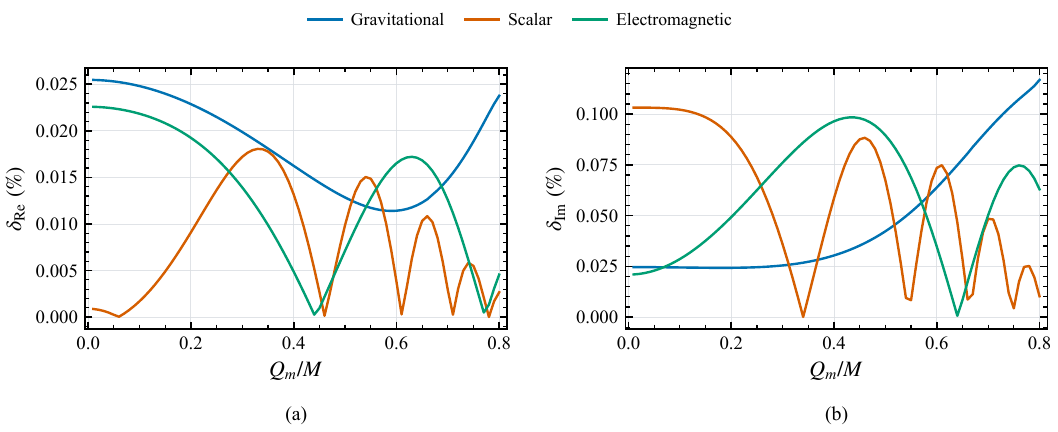}
\caption{
Same as Fig.~\ref{fig:updated-L2-errors-plus}, but for \(\eta=-1\) with
\(0.01\leq Q\leq0.80\).  The largest value occurs on the
gravitational-led \(|\operatorname{Im}(M\omega)|\) component at the
endpoint, where the background is closest to extremality.}
\label{fig:updated-L2-errors-minus}
\end{figure}

Before turning to the complex plane, we quantify the accuracy of the curves
just discussed.  Figs.~\ref{fig:updated-L1-errors-plus}--\ref{fig:updated-L2-errors-minus}
show the componentwise DI-PS relative differences for the four scans of
Figs.~\ref{fig:updated-L1-plus}--\ref{fig:updated-L2-minus}.  Within each
error panel, all fields of a fixed \(\ell\) and coupling sign are plotted
together with the real- and imaginary-part errors separated.

The purpose of displaying 
Figs.~\ref{fig:updated-L1-errors-plus}--\ref{fig:updated-L2-errors-minus}
is to establish the reliability of the frequency curves of
Figs.~\ref{fig:updated-L1-plus}--\ref{fig:updated-L2-minus}, on which the
preceding analysis rests.  Because direct integration and the
pseudospectral method are formulated independently (Sec.~\ref{sec4}),
their agreement along each scan is a cross-validation stronger than any
internal convergence test.  The agreement is close.  Every componentwise
relative difference stays sub-percent, with the largest values on the
\(\ell=1\), \(\eta=+1\) electromagnetic-led branch (\(0.4308\%\) and
\(0.4057\%\) in the real and imaginary parts near the upper end of the
interval) and with the constrained \(\ell=2\) system at or below
\(0.1169\%\).  The differences generally grow toward the endpoints, where
the background is most strongly deformed, and their smooth dependence on
\(Q\), without isolated spikes. It confirms that both methods track the
same discrete modes rather than occasionally switching branches.

In summary, the two independently formulated methods agree to sub-percent accuracy over the whole sampled domain, the residuals reflecting only their independent discretizations.  The spectral trends identified in this subsection therefore rest on firm numerical ground and we now follow the same branches into the complex-frequency plane.

\subsection{Complex QNM frequency trajectories}
\label{subsec:updated-complex-trajectories}

\begin{figure}[H]
\centering
\includegraphics[width=0.96\linewidth]{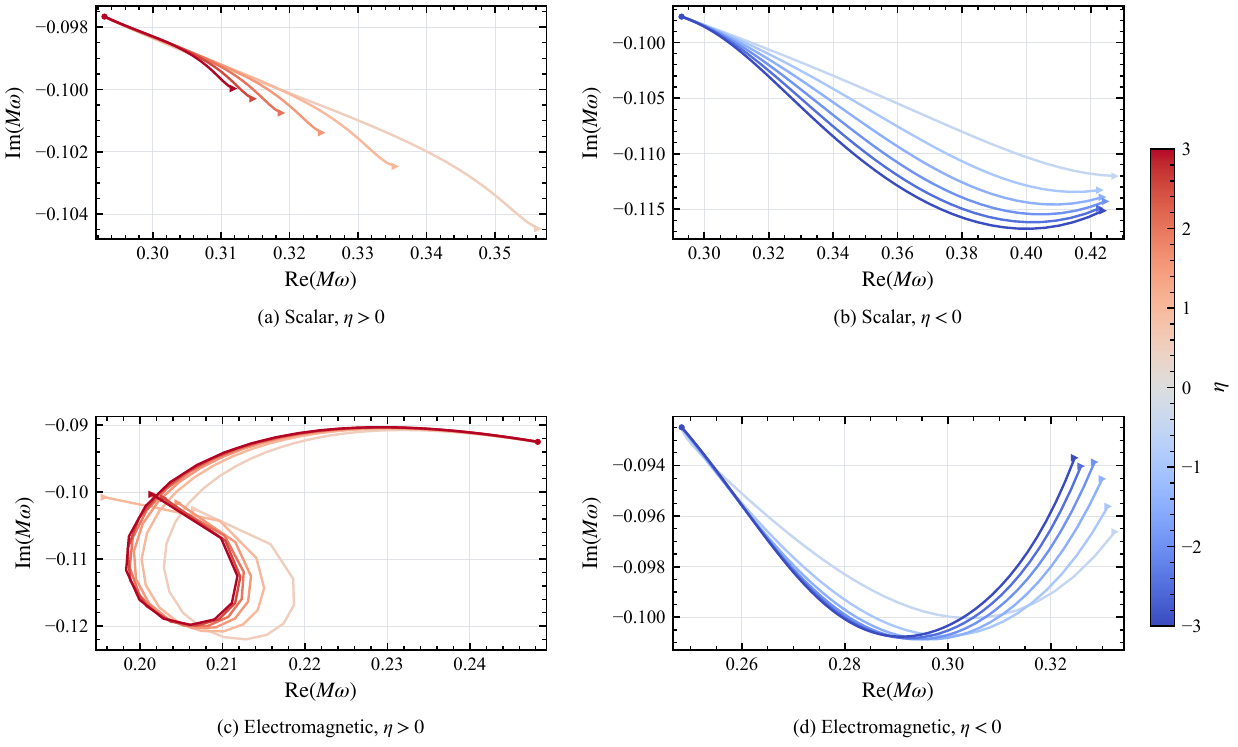}
\caption{
Pseudospectral complex-frequency trajectories of the \(\ell=1\) fundamental
modes (\(M=1\), \(n=0\), \(|\eta|\in\{0.5,1,1.5,2,2.5,3\}\)), the charge
running from \(Q=0.01\) up to near the critical value of
Table~\ref{tab:critical-charges}.  Rows: scalar-led and
electromagnetic-led sectors.  Columns: positive and negative \(\eta\).
Arrows indicate increasing \(Q\), and color encodes the coupling.
The loops of the electromagnetic-led trajectories represent simultaneous
reversals of the frequency and damping trends rather than a discontinuity
of the tracked mode.}
\label{fig:updated-complex-L1}
\end{figure}

\begin{figure}[H]
\centering
\includegraphics[width=0.96\linewidth]{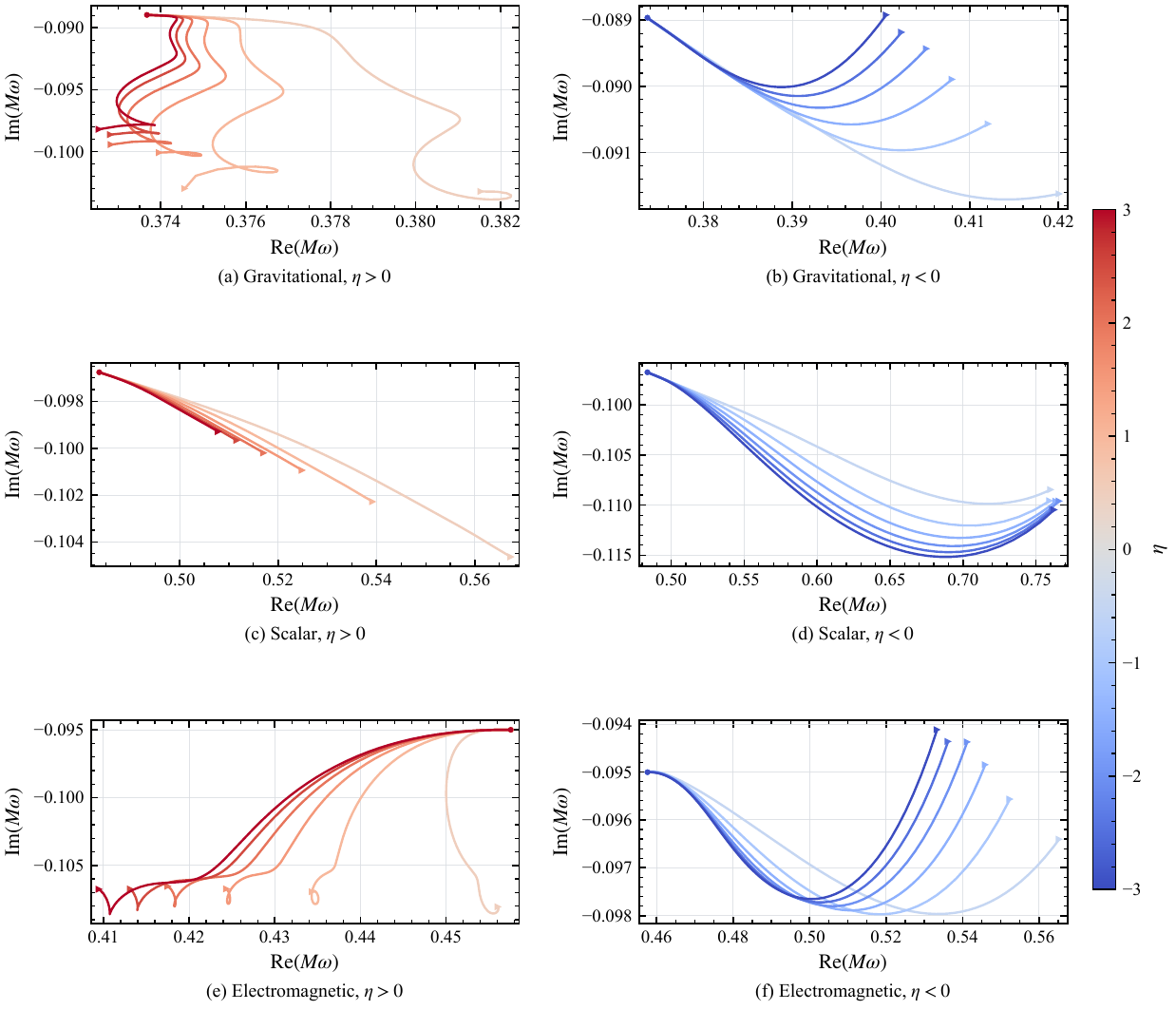}
\caption{
Pseudospectral complex-frequency trajectories of the \(\ell=2\) fundamental
modes (\(M=1\), \(n=0\), \(|\eta|\in\{0.5,1,1.5,2,2.5,3\}\)), the charge
running from \(Q=0.01\) up to near the critical value of
Table~\ref{tab:critical-charges}.  Rows: gravitational-led, scalar-led,
and electromagnetic-led sectors. Columns: positive and negative
\(\eta\).  Arrows indicate increasing \(Q\), and color encodes the
coupling. Separating the three sectors avoids a misleading comparison of
branches with different natural frequency scales.}
\label{fig:updated-complex-L2}
\end{figure}

Each charge scan of
Sec.~\ref{subsec:updated-di-ps-comparison} condenses into a single
trajectory when the two frequency components are plotted jointly in the
complex plane, with abscissa \(\operatorname{Re}(M\omega)\) and ordinate
\(\operatorname{Im}(M\omega)\).
The \(\ell=1\) trajectories are shown in Fig.~\ref{fig:updated-complex-L1}
and the \(\ell=2\) trajectories in Fig.~\ref{fig:updated-complex-L2}.  
As the curves of Figs.~\ref{fig:updated-L1-plus}--\ref{fig:updated-L2-minus} are shown,
they are computed with the pseudospectral method.  The panels are
separated by perturbation sector because the gravitational, scalar, and
electromagnetic branches occupy different regions of the complex plane.
Within a panel, the six curves correspond to
\(|\eta|=0.5,1,1.5,2,2.5,3\), the sign of \(\eta\) being given by the
column.  The endpoint trimming described in
Sec.~\ref{subsec:updated-data-selection} is applied before plotting.

In the complex plane, the spectral trends of
Sec.~\ref{subsec:updated-di-ps-comparison} acquire a geometric form.  The
scalar-led trajectories move monotonically toward larger real frequency,
their ordered family showing that increasing \(|\eta|\) changes the rate
of spectral deformation rather than the branch identity.  The
electromagnetic-led trajectories bend much more strongly and form loops
for \(\eta>0\)---the complex-plane image of the simultaneous extrema of
Figs.~\ref{fig:updated-L1-plus}--\ref{fig:updated-L2-minus}---while the
\(\ell=2\) gravitational-led family stays compact for \(\eta>0\) and fans
out for \(\eta<0\), reflecting its sensitivity both to the modified
background and to the coupling with the matter fields.  All trajectories
remain in the lower half of the complex plane. All fundamental modes are
damped over the whole resolved domain,  supporting that the dEH black hole is stable against $\ell=1,2$ mode-perturbations.

We note that the scans select \(Q=Q_m/M\geq0\).  For the coupled system considered here, since the
field equations are invariant under \(Q_m\to-Q_m\) together with
\(u_4\to-u_4\),  the QNM frequencies are even in the magnetic
charge.

\section{Conclusions and Discussions}
\label{sec6}

In this paper we formulated and solved the coupled QNM 
problem of the magnetically charged dEH black
hole for the even-parity gravitational, odd-parity electromagnetic,
and dilaton perturbations in the EEHd theory.  The monopole of the $\ell=0$ sector reduces to the single
dilaton equation studied previously~\cite{Li:2026gqi}, while the dipole
sector ($\ell=1$) couples the dilaton to the axial electromagnetic amplitude
through the magnetic charge.  However,  the quadrupole of the $\ell=2$ sector yields a
constrained five-mode system,  where the Einstein constraint must be
retained to exclude spurious modes.  All calculations were done in the ghost-free, horizon-bearing region of the \((\zeta, Q)\) parameter space with dilaton coupling difference $\zeta=\alpha-\beta$ and dimensionless charge $Q=Q_m/M$.

The fundamental frequencies of QNMs  were computed by introducing  two independent methods:
direct integration with Frobenius series at both boundaries and
matching at a matching radius, and Chebyshev pseudospectral
collocation of the compactified  boundary-factorized system with the
Einstein constraint imposed as a bordered row.  On the overlapping
scans with  \(\eta=\zeta/M^2=\pm1\) the two methods track the same branches and agree
closely.  The largest componentwise discrepancy over all common points
and sectors is \(0.4308\%\), while typical (median) differences are a
few times smaller.   This agreement guarantees both the numerical
solution and the mode identification based on frequency continuity and
eigenvector overlap.

Three important conclusions are as follows.  Firstly, every resolved fundamental branch of the \(\ell=1\) and \(\ell=2\) systems remains damped throughout the admissible domain, supporting the stable dEH black hole,  although this is not a perfect proof of the full linear stability.  Secondly, at finite charge each eigenfunction is composed of a hybrid mode (gravitational, dilaton, and electromagnetic modes), so a perturbation excited in one channel might radiate into the others. Relative to the gravitational–dilaton perturbation system without electromagnetic perturbation~\cite{Li:2026gqi}, the fundamental frequencies  are shifted by
up to tens of percent at high charge
(see Appendix~\ref{app:previous-reduced-comparison}).  Thirdly,
the spectral response is asymmetric in the sign of
\(\zeta\)---stiff in frequency and soft in damping for
\(\zeta>0\), reverse for \(\zeta<0\) with matter-led branches
drifting by up to tens of percent---so precise  ringdown measurements
could  constrain both \(Q\) and the sign of \(\zeta\).

Taken all together, our results established the polar-led fundamental
spectrum of the dEH black hole on firm numerical ground and identified 
the dimensionless  charge \(Q\) and the sign of the dimensionless coupling \(\eta\)
as its two control parameters.  These  might provide the baseline against which
Type I (the remaining sector of the full perturbations) can be analyzed.

 \vspace{1cm}
{\bf Acknowledgments}
 \vspace{1cm}

We would like to thank Ming Zhang and Chao-Ming Zhang for helpful discussions. We gratefully acknowledge support by the National Natural Science Foundation of China (NNSFC) (Grant No.12365009), Jiangxi Provincial
Natural Science Foundation (Grant No. 20262BAC240347) and the National Research Foundation of Korea (No.RS-2022-NR069013).

\appendix
\section{Dimensionless perturbation equations}
\label{appendix-A}

This appendix records the dimensionless form of the \(\ell=1\) and \(\ell=2\)
perturbation equations used in the pseudospectral method calculation.
Let us introduce five dimensionless quantities
\begin{align}
 Q&=\frac{Q_m}{M}, & \eta&=\frac{\zeta}{M^2}, & \Omega&=M\omega,\label{eq:dimensionless-map}\\
 r&=M\sqrt{y(y-Q^2)}, &
  y&=\frac{\rho}{M}
 =\frac{Q^2+\sqrt{Q^4+4r^2/M^2}}{2}.\nonumber
\end{align}
For $y>Q^2$, the chain rule gives
\begin{align}
 \frac{ d}{ d r}
 &=\frac{2\sqrt{y(y-Q^2)}}{M(2y-Q^2)}\frac{ d}{ d y},
 \label{eq:first-derivative-map}\\
 \frac{ d^2}{ d r^2}
 &=\frac{4y(y-Q^2)}{M^2(2y-Q^2)^2}\frac{ d^2}{ d y^2}
 +\frac{2Q^4}{M^2(2y-Q^2)^3}\frac{ d}{ d y}.
 \label{eq:second-derivative-map}
\end{align}
Also, we have 
\begin{equation}
 A(r)=A_y(y),\quad B(r)=B_y(y),\quad \phi(r)=\bar\phi(y),
 \qquad \phi_1(r)=\Phi_1(y),\qquad
 u_4(r)=\frac{r}{M}U_4(y).
 \label{eq:l1-amplitude-map}
\end{equation}

The dimensionless equations are used for the amplitude normalizations of the
symbolic derivation with \(M=1\).  In
particular, \(\Phi_1\) and \(U_4\) are not new modes and the
quadrupole modes of \(H_0,K,R_h\) follow the same convention as the
coefficient definitions below.
For the dHE black hole background, one has
\begin{align}
 A_y(y)&=1-\frac{2}{y}-\frac{2\eta Q^4}{y^3(y-Q^2)^3},
 \label{eq:Ay-background}\\
 B_y(y)&=A_y(y)\frac{(2y-Q^2)^2}{4y(y-Q^2)},
 \label{eq:By-background}\\
 \bar\phi(y)&=-\frac12\ln\left(\frac{y-Q^2}{y}\right).
 \label{eq:phiy-background}
\end{align}
To obtain  the dimensionless equations, let us define
\begin{equation}
 s=Q^2-y,\qquad p=Q^2-2y,\qquad
 \mathcal C=3+4 e^{2\bar\phi}+3 e^{4\bar\phi},
 \qquad \mathcal W=s^2y^2-2\eta Q^2\mathcal C.
 \label{eq:compact-short-def}
\end{equation}
Here $s$ and $p$ are related to the original radial coordinate $r$ through $y=r/M$.
For the outside region of the horizon, one finds that  \(s=Q^2-y<0\) and \(p=Q^2-2y<0\).  The signs are useful
when comparing the dimensionless expressions with the physical-radius equations.
The remaining dimensionless coefficients  are given by 
\begin{align}
 \mathcal X={}&y^2Q^4-2Q^2\left(y^3+\eta\mathcal C\right)+y^4,
 \label{eq:X-def}\\
 \mathcal Y={}&y^2Q^4-2Q^2
 \left[y^3-3\eta\left( e^{4\bar\phi}-1\right)\right]+y^4,
 \label{eq:Y-def}\\
 \mathcal Z={}&3\eta Q^4 e^{2\bar\phi}+y^3s^3
 -Q^2 e^{-2\bar\phi}
 \left[Q^4y^2-Q^2(3\eta+2y^3)+y^4\right].
 \label{eq:Z-def}
\end{align}
Here and below, $\bar\phi$, $A_y$, $B_y$, and the calligraphic factors in the
dimensionless equation are evaluated at fixed  $y$.
Furthermore, we note that 
\(\mathcal X=y^2(y-Q^2)^2-2\eta Q^2\mathcal C\) coincides with
\(\mathcal W\).  Both symbols are retained because the scalar and
electromagnetic equations below use them in different combinations.

\subsection{The \(\ell=1\)-mode equation}

Substituting Eqs.~\eqref{eq:dimensionless-map}--\eqref{eq:l1-amplitude-map}
and normalizing the coefficient of \(\Phi_1''\), the scalar
equation \eqref{eq:l1-scal2} becomes the dimensionless equation
\begin{align}
0={}&\Phi_1''
+\frac12\left(\frac{A_y'}{A_y}+\frac{B_y'}{B_y}
+\frac{Q^4}{y(Q^4-3Q^2y+2y^2)}\right)\Phi_1'
\nonumber\\
&+\Bigg[\frac{pA_y'}{4yA_y(y-Q^2)}
+\frac{A_yp^2}{y^2B_y(y-Q^2)^2}
+\frac{\Omega^2p^2}{4yA_yB_y(y-Q^2)}
+\frac{pB_y'}{4yB_y(y-Q^2)}
\nonumber\\
&\hspace{1.2cm}
+\frac{p^2\mathcal Z}{2y^5B_y s^4(y-Q^2)}
+\frac{p^2}{y^2s(y-Q^2)}\Bigg]\Phi_1
\nonumber\\
&+\frac{Qp e^{-2\bar\phi}}
 {y^4B_y s^4}
 \left[2yB_y s\bar\phi'\mathcal X+p\mathcal Y\right]U_4
+\frac{4Q e^{-2\bar\phi}\bar\phi'\mathcal X}{y^2s^2}U_4'.
\label{eq:l1-scaly1}
\end{align}
Similarly, the electromagnetic equation
\eqref{eq:l1-maxeq2} becomes the dimensionless equation
\eqref{eq:l1-maxeqy1}:
\begin{align}
0={}&U_4''+\mathcal A_U U_4'+\mathcal B_U U_4
-\frac{Qp^2\mathcal Y}{2s^3y^3\mathcal X B_y}\Phi_1,
\label{eq:l1-maxeqy1}\\
\mathcal A_U={}&-\frac{1}{2y(y-Q^2)p}
\Bigg\{Q^4+p^2\Bigg[
 \frac{4s^2y^2}{\mathcal X}-2
\nonumber\\
&\hspace{2.7cm}
+\frac{sy}{p}\left(
 \frac{A_y'}{A_y}+\frac{B_y'}{B_y}
 -\frac{4\mathcal Y\bar\phi'}{\mathcal X}\right)
\Bigg]\Bigg\},
\label{eq:l1-AU}\\
\mathcal B_U={}&\frac14\Bigg[
-\frac{2\mathcal C p\eta Q^2B_y'}{sy\mathcal W B_y}
-\frac{24( e^{4\bar\phi}-1)p\eta Q^2\bar\phi'}{sy\mathcal W}
+\frac{8\mathcal C p^2\eta Q^2}{s^2y^2\mathcal W}
\nonumber\\
&\hspace{1.0cm}
+\frac{12\mathcal C p^2\eta Q^2}{s^2y^2\mathcal W B_y}
+\frac{pA_y'}{syA_y}
\nonumber\\
&\hspace{1.0cm}
+\frac{(2y-Q^2)
 \left[2Q^2-4y+y(y-Q^2)B_y'+4syB_y\bar\phi'\right]}
 {\mathcal W B_y}\Bigg]
-\frac{p^2\Omega^2}{4syA_yB_y}.
\label{eq:l1-BU}
\end{align}
We note that equations~\eqref{eq:l1-scaly1} and \eqref{eq:l1-maxeqy1} are used for the numerical computation, written in terms of 
dimensionless exterior radius \(y=r/M\) with $y>Q^2$.

\subsection{The \(\ell=2\)-mode equation}
\label{subsec:l2-publication-equations}

For the \(\ell=2\) sector, the five amplitudes are included as 
\[
 \boldsymbol{\Psi}_2(y)=
 \bigl(K_p(y),R_h(y),H_0(y),\Phi_1(y),U_4(y)\bigr)^{\mathsf T},
 \qquad y>Q^2 .
\]
A prime in the following equations denotes \(d/dy\). To avoid a reduced font size, we first
isolate the repeated polynomial factors.  Let us define
\begin{align}
 \Delta&=y-Q^2, & P&=2y-Q^2, & \Lambda&=\ell(\ell+1),
 \label{eq:l2-readable-basic}\\
 \mathcal X_3&=Q^4y^2-2Q^2\left(y^3+3\eta\mathcal C\right)+y^4,
 \nonumber\\
 \mathcal T&=Q^4y^2-2Q^2\left[y^3+\eta
   \left(3+2e^{2\bar\phi}\right)\right]+y^4 .
 \label{eq:l2-readable-polynomials}
\end{align}
Here \(\mathcal C\), \(\mathcal W\), \(\mathcal X\), and \(\mathcal Y\)
are already defined in Eqs.~\eqref{eq:compact-short-def}--\eqref{eq:Y-def}.  All
background and coefficient functions below are evaluated at fixed \(y\),
so their arguments are suppressed to improve legibility.

We introduce  three combined coefficients 
\begin{align}
 \mathcal G_K={}&-2y^3\Delta^3(A_y-2B_y)
 +\frac{4y^4\Delta^4}{P}B_y'
 +(\Lambda-2)y^3\Delta^3
 \nonumber\\
 &-2\eta Q^4e^{-2\bar\phi}\mathcal C
 +2Q^2y^2\Delta^2e^{-2\bar\phi},
 \label{eq:l2-GK}\\
 \mathcal G_H={}&\frac{2y^4\Omega^2\Delta^4}{A_y}
 +2y^3A_y\Delta^3
 -y^3\Delta^3\left(\frac{4y\Delta B_y'}{P}+\Lambda-2\right)
 \nonumber\\
 &-4y^3B_y\Delta^3+2\eta Q^4e^{-2\bar\phi}\mathcal C
 -2Q^2y^2\Delta^2e^{-2\bar\phi},
 \label{eq:l2-GH}\\
 \mathcal G_R={}&2y^3A_y\Delta^3e^{2\bar\phi}
 -\frac{y^3\Delta^3e^{2\bar\phi}}{P}
 \left[4y\Delta B_y'+(\Lambda-2)P\right]
 \nonumber\\
 &-4y^3B_y\Delta^3e^{2\bar\phi}
 +2\eta Q^4\mathcal C-2Q^2y^2\Delta^2 .
 \label{eq:l2-GR}
\end{align}
Then, the three first-order gravitational equations take
the readable forms as 
\begin{align}
0={}&K_p'+\frac{2PA_y-y\Delta A_y'}{2yA_y\Delta}K_p
-\frac{P}{2y\Delta}H_0+\frac{2\bar\phi'}{y\Delta}\Phi_1
-\frac{iP\mathcal G_K}{4y^5\Delta^5}R_h,
\label{eq:l2-publication-eqKy1}\\
0={}&R_h'+\frac{B_yA_y'+A_yB_y'}{2A_yB_y}R_h
+\frac{iP}{2B_y}(H_0+K_p)
-\frac{2iQP e^{-2\bar\phi}\mathcal X}{y^3B_y\Delta^3}U_4,
\label{eq:l2-publication-eqRy1}\\
0={}&H_0'+\frac{A_y'}{A_y}H_0
+\frac{PA_y-y\Delta A_y'}{2yA_y\Delta}K_p
+\frac{iP\mathcal G_H}{4y^5\Delta^5}R_h
-\frac{2\bar\phi'}{y\Delta}\Phi_1
\nonumber\\
&+\frac{4Qe^{-2\bar\phi}\mathcal X}{y^3\Delta^3}U_4'
+\frac{2QPe^{-2\bar\phi}\mathcal X}{y^4\Delta^4}U_4 .
\label{eq:l2-publication-eqH0y1}
\end{align}
These equations show  how the scalar and electromagnetic
amplitudes are sources for the gravitational sector.  In particular, all mixing terms
vanish in the limit of  \(Q\to0\), providing  an immediate check for the
neutral limit.

The remaining Einstein equation is algebraic in the metric
part and first order in the matter part.  We rewrite it as
\begin{equation}
 \mathcal C_R R_h+\mathcal C_H H_0+\mathcal C_K K_p
 +\mathcal C_\Phi\Phi_1+\mathcal C_{\Phi'}\Phi_1'
 +\mathcal C_U U_4+\mathcal C_{U'}U_4'=0,
 \label{eq:l2-publication-eqcoy1}
\end{equation}
whose coefficients are given by 
\begin{align}
\mathcal C_R={}&-iyB_y\Delta
 \left(2y^3\Omega^2\Delta^3e^{2\bar\phi}
 +\frac{A_y'}{P}\mathcal G_R\right),
\label{eq:l2-constraint-CR}\\
\mathcal C_H={}&y\left\{
 \frac{6y^4B_y\Delta^5e^{2\bar\phi}A_y'}{P}
 +A_y\Delta\left[(\Lambda-2)y^3\Delta^3e^{2\bar\phi}
 -2\eta Q^4\mathcal C+2Q^2y^2\Delta^2\right]\right\},
\label{eq:l2-constraint-CH}\\
\mathcal C_K={}&y\left\{
 \frac{2y^5B_y\Delta^6e^{2\bar\phi}A_y'^2}{A_yP^2}
 -\frac{2y^4B_y\Delta^5e^{2\bar\phi}A_y'}{P}
 +2y^4\Omega^2\Delta^5e^{2\bar\phi}
 \right.\nonumber\\
&\left.\hspace{1.4cm}
 +A_y\Delta\left[-(\Lambda-2)y^3\Delta^3e^{2\bar\phi}
 +8\eta Q^4\mathcal C-4Q^2y^2\Delta^2\right]\right\},
\label{eq:l2-constraint-CK}\\
\mathcal C_\Phi={}&-4\left\{
 \frac{2y^5B_y\Delta^5e^{2\bar\phi}A_y'\bar\phi'}{P^2}
 +A_y\left[\frac{2y^4B_y\Delta^4e^{2\bar\phi}\bar\phi'}{P}
 \right.\right.\nonumber\\
&\left.\left.\hspace{4.2cm}
 +3\eta Q^4(e^{4\bar\phi}-1)+Q^2y^2\Delta^2\right]\right\},
\label{eq:l2-constraint-CPhi}\\
\mathcal C_{\Phi'}={}&-\frac{16y^5A_yB_y\Delta^5
 e^{2\bar\phi}\bar\phi'}{P^2},
\nonumber\\
\mathcal C_U={}&4QyA_y\Delta(2B_y+\Lambda)\mathcal W,
\label{eq:l2-constraint-matter1}\\
\mathcal C_{U'}={}&\frac{16Qy^2A_yB_y\Delta^2\mathcal X}{P}.
\label{eq:l2-constraint-matter2}
\end{align}
Note that equation~\eqref{eq:l2-publication-eqcoy1} is not an extra propagating
equation.  It enforces a consistency among the five amplitudes and it is 
valuable numerically because a mode with a small eigenvalue residual can
 be rejected if its independent constraint residual is large.

Finally, the dilaton  and electromagnetic equations take the forms 
\begin{align}
0={}&\Phi_1''+\mathcal A_\Phi\Phi_1'
+\mathcal V_\Phi\Phi_1+\mathcal S_{\Phi K}K_p
+\mathcal S_{\Phi U}U_4+\mathcal S_{\Phi U'}U_4',
\label{eq:l2-publication-scaly1}\\
0={}&U_4''+\mathcal A_U U_4'+\mathcal V_U U_4
+\mathcal S_{UK}K_p+\mathcal S_{U\Phi}\Phi_1
\label{eq:l2-publication-maxeqy1}
\end{align}
with coefficients
\begin{align}
\mathcal A_\Phi={}&\frac12\left(
 \frac{A_y'}{A_y}+\frac{B_y'}{B_y}+\frac{Q^4}{y\Delta P}\right),
\label{eq:l2-scalar-A}\\
\mathcal V_\Phi={}&\frac{P^2}{8y\Delta}\left\{
 -\frac{2(PA_y'+16yA_y\Delta\bar\phi'^2)}{A_yP^2}
 +\frac{2\Omega^2}{A_yB_y}
 \right.\nonumber\\
&\left.\quad
 +\frac{e^{-2\bar\phi}}{y^4B_y\Delta^4}\left[
 -\frac{2y^3\Delta^3e^{2\bar\phi}}{P}
 (y\Delta B_y'+\Lambda P)
 +12\eta Q^4(e^{4\bar\phi}+1)-4Q^2y^2\Delta^2
 \right]\right\},
\label{eq:l2-scalar-V}\\
\mathcal S_{\Phi K}={}&-\frac{Q^2P^2e^{-2\bar\phi}\mathcal Y}
 {2y^4B_y\Delta^4},
\nonumber\\
\mathcal S_{\Phi U'}={}&\frac{4Qe^{-2\bar\phi}
 \bar\phi'\mathcal X}{y^2\Delta^2},
\label{eq:l2-scalar-sources1}\\
\mathcal S_{\Phi U}={}&\frac{QP^2e^{-2\bar\phi}}
 {2y^4B_y\Delta^4}\left\{
 y^2\Delta^2\left(\frac{4yB_y\Delta\bar\phi'}{P}+\Lambda\right)
 \right.\nonumber\\
&\left.\hspace{2.5cm}
 +2\eta Q^2\left[3\Lambda(e^{4\bar\phi}-1)
 -\frac{4yB_y\Delta\mathcal C\bar\phi'}{P}\right]\right\},
\label{eq:l2-scalar-sources2}\\
\mathcal A_U={}&\frac12\left\{
 \frac{A_y'}{A_y}+\frac{B_y'}{B_y}-\frac4P-\frac1\Delta
 +\frac{4y\Delta P-8\bar\phi'\mathcal T}{\mathcal X}
 +4\bar\phi'-\frac1y\right\},
\label{eq:l2-em-A}\\
\mathcal V_U={}&\frac{P^2}{4y^2\Delta^2}\left\{
 \frac1P\left[y\Delta\left(\frac{A_y'}{A_y}+\frac{B_y'}{B_y}\right)
 +\frac{8\eta Q^2P\mathcal C-4y\Delta\bar\phi'\mathcal Y}
 {\mathcal X}\right]
 \right.\nonumber\\
&\left.\hspace{1.8cm}
 +\frac1{B_y}\left[\frac{y\Omega^2\Delta}{A_y}
 -\frac{3\Lambda}{1+2y^2\Delta^2/\mathcal X_3}\right]\right\},
\label{eq:l2-em-V}\\
\mathcal S_{UK}={}&\frac{QP^2\mathcal X_3}
 {4y^2B_y\Delta^2\mathcal X},
\nonumber\\
\mathcal S_{U\Phi}={}&\frac{QP^2\mathcal Y}
 {2y^3B_y\Delta^3\mathcal X}.
\label{eq:l2-em-sources}
\end{align}
The above representation is a condensed rewrite of the direct
symbolic output.  It preserves the differential order, the constraint,
and the inter-sector couplings.  A symbolic substitution check of the
printed coefficient functions is required  before using them in an
independent implementation.
It is worth noting that expressions~\eqref{eq:l2-publication-eqKy1}--\eqref{eq:l2-publication-maxeqy1}
include  six equations of the quadrupolar ($\ell=2$) system: three first-order
gravitational equations, one Einstein constraint, and two second-order
matter equations.  Furthermore, the constraint is not discarded in the numerical
problem.  It enters as a bordered row of the pencil and is also evaluated
independently on the collocation grid.

\subsection{Compactification for the numerical solution}

Carrying out the pseudospectral discretization, both systems are mapped from the
unbounded exterior \(y>Q^2\) onto a finite interval through
\begin{equation}
 x=1-\frac{y_h}{y},\qquad y=\frac{y_h}{1-x},\qquad x\in[0,1)
 \label{eq:finite-interval-map}
\end{equation}
and with constraints from the metric at the horizon
\begin{equation}
\eta = \frac{(y_h-2) {y_h}^2 \left(y_h-Q^2\right)^3}{2 Q^4}.
 \label{eq:eta-constraints}
\end{equation}
This  map is written as the implementation form in terms of 
\(y\).  It is a separate reparametrization from the generic \(r\)-space map
in Eq.~\eqref{eq:ps-compact-map}.  They describe the same endpoints, but
their finite-resolution node distributions do not need to coincide.
Here, \(y_h\) denotes the outer horizon.   Hence, we have relations
\begin{equation}
 \frac{ d}{ d y}=\frac{(1-x)^2}{y_h}\frac{ d}{ d x},
 \qquad
 \frac{ d^2}{ d y^2}=
 \frac{(1-x)^4}{y_h^2}\frac{ d^2}{ d x^2}
 -\frac{2(1-x)^3}{y_h^2}\frac{ d}{ d x}.
 \label{eq:y-to-x-derivatives}
\end{equation}
Applying this compactification to the \(l=1\) and \(l=2\) systems leads to
the finite Chebyshev systems used for the QNM calculation.

\section{Comparison with the gravitational-dilaton system}
\label{app:previous-reduced-comparison}

\begin{figure}[H]
\centering
\subfigure[~$\operatorname{Re}(M\omega)$]{
\label{fig:l=1scRecomparison}
\includegraphics[height=1.9in]{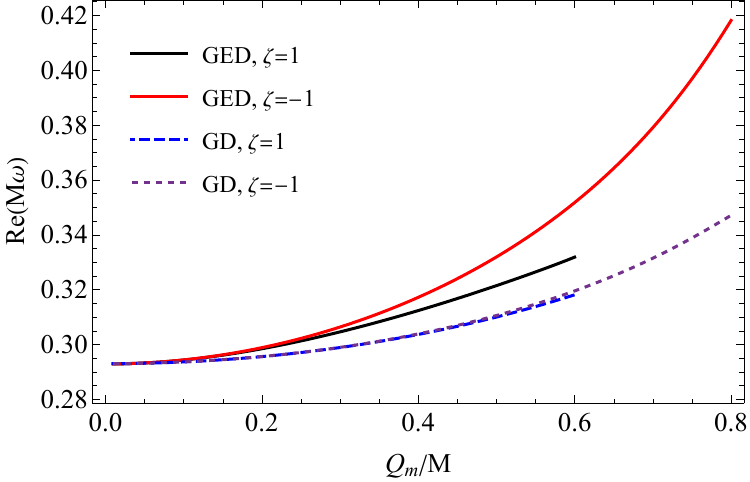}}
\quad
\subfigure[~$\operatorname{Im}(M\omega)$]{
\label{fig:l=1scImcomparison}
\includegraphics[height=1.9in]{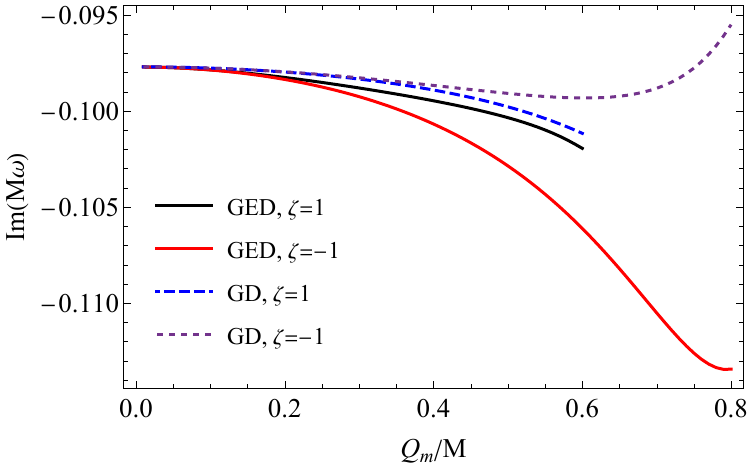}}
\caption{QNM frequencies as functions of $Q_m/M$ for 
 fundamental scalar-led \(\ell=1\) mode with \(\zeta=\pm 1\) \((M=1\), \(n=0)\). Comparison between gravitational-electromagnetic-dilaton perturbation system (GED) and gravitational–dilaton perturbation system (GD) of Ref.~\cite{Li:2026gqi}. }
\label{fig:l=1sccomparison}
\end{figure}

\begin{figure}[H]
\centering
\subfigure[~$\operatorname{Re}(M\omega)$]{
\label{fig:l=2scRecomparison}
\includegraphics[height=1.9in]{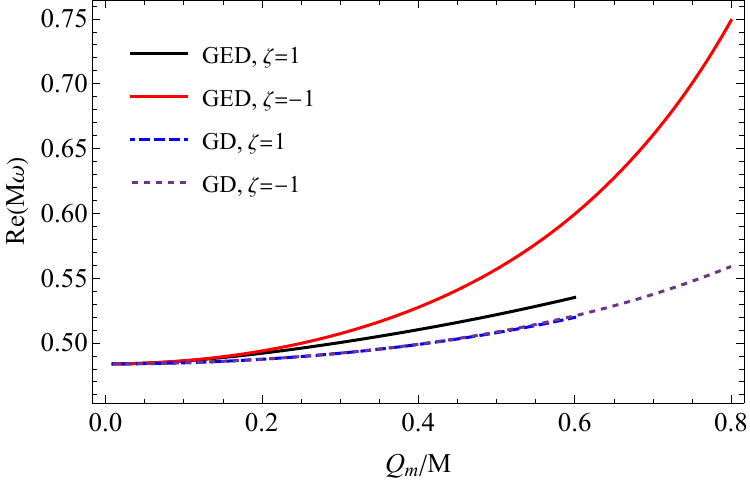}}
\quad
\subfigure[~$\operatorname{Im}(M\omega)$]{
\label{fig:l=2scImcomparison}
\includegraphics[height=1.9in]{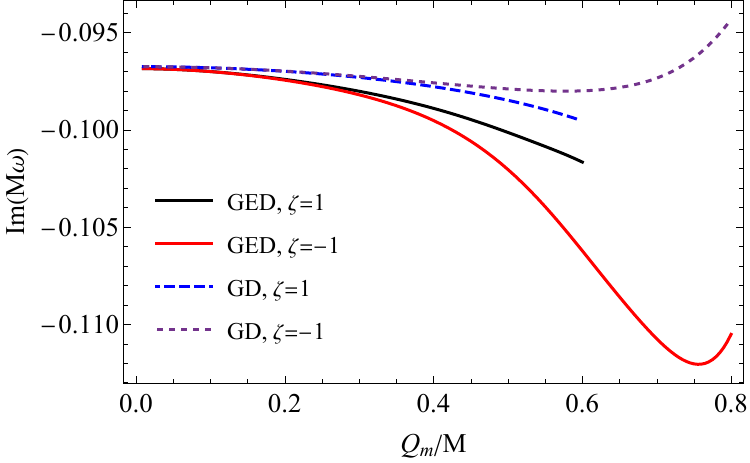}}
\caption{
 QNM frequencies as functions of $Q_m/M$ for the scalar-led \(\ell=2\) mode.}
\label{fig:l=2sccomparison}
\end{figure}

\begin{figure}[H]
\centering
\subfigure[~$\operatorname{Re}(M\omega)$]{
\label{fig:l=2grRecomparison}
\includegraphics[height=1.9in]{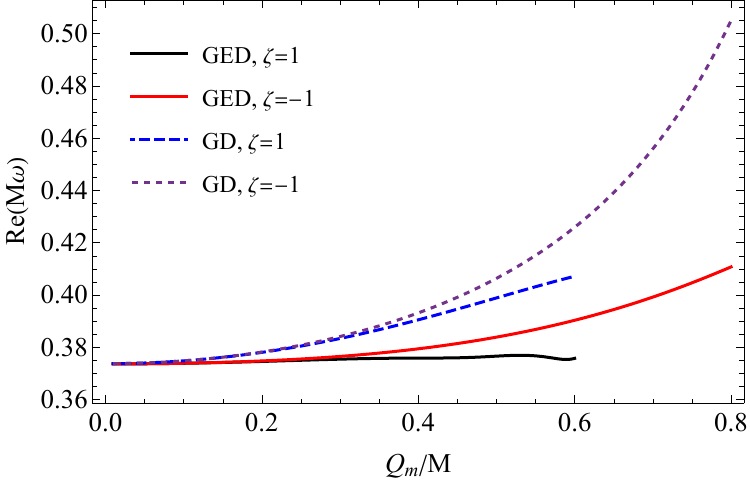}}
\quad
\subfigure[~$\operatorname{Im}(M\omega)$]{
\label{fig:l=2grImcomparison}
\includegraphics[height=1.9in]{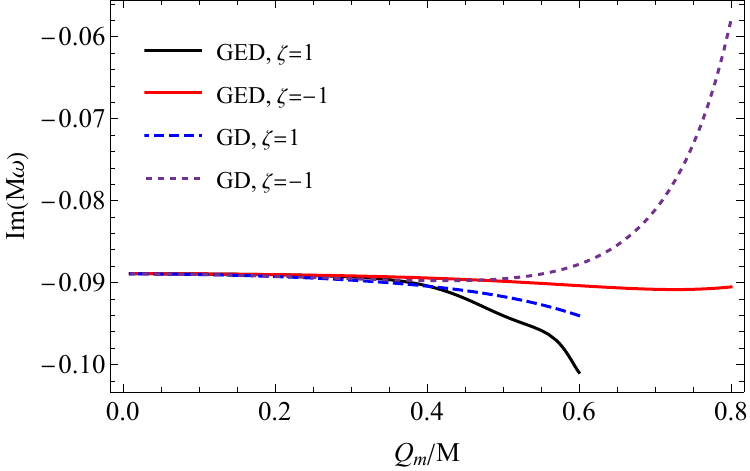}}
\caption{
 QNM frequencies as functions of $Q_m/M$ for  the gravitational-led \(\ell=2\) mode.}
\label{fig:l=2grcomparison}
\end{figure}

We note that Ref.~\cite{Li:2026gqi} has retained a dilaton-only
dipole and a gravitational-dilaton quadrupole, omitting the
axial electromagnetic perturbation.   The comparison therefore serves
two purposes. Let us verify that both formulations share the Schwarzschild
limit and quantify the spectral shifts which are produced by 
electromagnetic sector in the strong-field regime.  In
Figs.~\ref{fig:l=1sccomparison}--\ref{fig:l=2grcomparison}, we compare the direct-integration QNM frequencies of the gravitational–dilaton perturbation system with those of the gravitational-electromagnetic-dilaton perturbation system studied here.  The coupling
constant of Ref.~\cite{Li:2026gqi} is denoted \(\epsilon\)  as
\(\epsilon=\zeta\) and the charge is \(Q=Q_m/M\) with
\(M=1\).  In the legends, GD labels the gravitational–dilaton perturbation and GED the gravitational-electromagnetic-dilaton perturbation of the present work.  It is clear that the common Schwarzschild limit fixes mode pairing.  The previous
\(\ell=1\) dilaton mode corresponds to the current scalar-led dipole,
and the previous \(\ell=2\) gravitational-dilaton mode corresponds to the
 current gravitational and scalar-led quadrupoles. These labels record ancestry rather than pure field content at finite charge. The
comparison covers \(0.01\leq Q\leq0.60\) for \(\zeta=+1\) and
\(0.01\leq Q\leq0.80\) for \(\zeta=-1\). These denote the charge intervals analyzed in Sec.~\ref{sec:updated-numerical-results}.

 It is worth noting that the ancestry-matched curves approach the same neutral limits
because the electromagnetic mixing terms vanish as \(Q\to0\), whereas their
separation generally becomes more pronounced toward the high-charge end.  To
quantify this apparent  change, we would like to  define a relevant quantity between these two formulations
\begin{equation}
 \Delta_{\rm form}(Q)=
 \frac{|M\omega_{\rm GED}^{\rm DI}(Q)-M\omega_{\rm GD}^{\rm DI}(Q)|}
 {|M\omega_{\rm GD}^{\rm DI}(Q)|}\times100\% .
 \label{eq:appendix-formulation-shift}
\end{equation}
At \(Q=0.01\), \(\Delta_{\rm form}\) lies between \(0.020\%\) and
\(0.033\%\) for six comparable branches.  At \(Q=0.60\) for
\(\zeta=+1\), it is \(4.12\%\), \(7.70\%\), and \(2.95\%\) for the
\(\ell=1\) scalar-led, \(\ell=2\) gravitational-led, and \(\ell=2\)
scalar-led modes, respectively.   At \(Q=0.80\) for \(\zeta=-1\), the
corresponding values are \(20.35\%\), \(19.62\%\), and \(33.61\%\).
Thus, the electromagnetic completion has little effect near the neutral limit
but it can produce substantial strong-field shifts, most prominently for the
\(\ell=2\), \(\zeta=-1\) scalar-led mode.

These offsets characterize the change in the perturbation operator due to the inclusion of nonlinear electromagnetic coupling. They are not direct-integration errors. We assess the numerical accuracy of the present three-field coupled GED system by an independent DI–PS comparison in Sec.~\ref{subsec:updated-di-ps-comparison}. Therefore, the comparison demonstrates that the axial electromagnetic amplitude, although absent from the gravitational–dilaton perturbation system, is regarded as a leading component of the strong-field spectrum of the dEH black hole.


\bibliographystyle{unsrtnat}
\bibliography{ref}

\end{document}